\documentclass{aa}  
\usepackage{gensymb}
\usepackage{graphicx}
\usepackage{txfonts}
\usepackage{lipsum}
\usepackage{subfig}         
\usepackage{lscape}             
\usepackage{placeins}           
\usepackage{tabularx}

\begin{document}

   \title{A stray light analysis for SO/PHI-HRT and an updated comparison of the inferred magnetic field with SDO/HMI}



   \author{J.~Sinjan\inst{1}\thanks{\hbox{Corresponding author: J. Sinjan} \hbox{\email{sinjan@mps.mpg.de}}}
        \and T.~L.~Riethm\"uller\inst{1}
        \and A.~Gandorfer\inst{1}
        \and A.~Feller\inst{1}
        \and D.~Calchetti\inst{1}
        \and F.~J.~Bail\'en\inst{2,3}
        \and J.~Hirzberger\inst{1}
        \and G.~Valori\inst{1}
        \and S.~K.~Solanki\inst{1}}

   \institute{Max-Planck-Institut f\"ur Sonnensystemforschung, Justus-von-Liebig-Weg 3,
         37077 G\"ottingen, Germany        
         \and 
         Instituto de Astrofísica de Andalucía (IAA-CSIC), Apartado de Correos 3004, 18080 Granada, Spain
         \and
         Spanish Space Solar Physics Consortium (S$^{3}$PC)}

   

   \date{Received August 30, 20XX}

 
  \abstract
   {The High Resolution Telescope of the Polarimetric and Helioseismic Imager on Solar Orbiter (SO/PHI-HRT) operates in an extreme observational environment, observing the Sun as close as $0.28$\;au. The high thermal load and large illuminating field puts high demands on the instrument in terms of both imaging performance and false light control.}
   {To characterise the amount of stray light (false light) within SO/PHI-HRT, apply a correction, and re-compare the data products with the Helioseismic and Magnetic Imager on the Solar Dynamics Observatory (SDO/HMI).} 
   {We analyse solar limb profiles and a Mercury transit to quantify the amount of stray light and add a correction term when partially reconstructing the SO/PHI-HRT images. For the comparison with SDO/HMI we use data from the 2023 March Solar Orbiter inferior conjunction and compare the magnetic fields on a pixel-by-pixel basis.}
   {Increased continuum intensity contrast in the quiet Sun, and darker intensity levels are found in strong magnetic features. Consequently, much stronger fields are inferred in these features. Comparing the stray light corrected data with that from the standard SDO/HMI data products results in a much closer agreement across all vector magnetic field components, particularly when the cadence and noise levels are identical. In most solar features, SO/PHI-HRT infers stronger fields than the SDO/HMI line-of-sight magnetograms. Compared to the vector magnetic field from SDO/HMI the two are very well aligned, with only slight differences in the strongest field regions (where $|\mathbf{B}|>1600$\;G or $|\mathbf{B_{\text{LOS}}}|>1300$\;G).}
   {}

   \keywords{Sun: magnetic fields -- Sun: photosphere -- instrumentation: high angular resolution -- space vehicles: instruments -- methods: data analysis}

   \maketitle


\section{Introduction}
The Polarimetric and Helioseismic Imager on Solar Orbiter \citep[SO/PHI, ][]{2020A&A...642A..11S,2020A&A...642A...1M} is a space-based solar magnetograph, which due to the orbit of Solar Orbiter, is the first to observe out of the Sun-Earth line, and over a wide range of distances to the Sun. It samples the \ion{Fe}{i}\,$6173.3$\,\AA\ absorption line and infers the photospheric vector magnetic field and line-of-sight velocity. The High Resolution Telescope of SO/PHI \citep[SO/PHI-HRT, ][]{2018SPIE10698E..4NG} performs routine observations of small portions of the Solar disk at close approaches to the Sun, while the Full Disk Telescope (SO/PHI-FDT) observes the entire disk, mostly outside the perihelion windows. 

The point spread function (PSF) describes the response of an optical instrument to a single point source of light. For an ideal optical telescope the PSF is the Airy function with the first minima at the diffraction limit. For space based solar telescopes, optical aberrations, and spacecraft jitter are the main sources of non-ideal imaging performance. 

An accurate characterization of the PSF for such telescopes is challenging. In the case of SO/PHI-HRT, the core of the point spread function is estimated by phase diversity wavefront analysis, which uses combinations of focused and defocused images to estimate the wavefront error \citep[][]{2022SPIE12180E..3FK,2023A&A...675A..61K,2024A&A...681A..58B}. The width of the PSF core is a measure of the spatial resolution. The wings of the PSF, corresponding to higher order spatial frequencies, lie outside the sensitivity range of phase diversity analysis \citep[][]{1983SPIE..351...56G,1985SPIE..528..202G}. The large spatial scale contribution to the wings of the PSF is commonly referred to as stray light. 

From a strict instrumental viewpoint however the term stray light (or false light) is used to describe any unwanted light that reaches the detector or image plane of a telescope. It is considered false as it can be light that does not come from the target object, e.g. as the result of unwanted reflections/scattering from dust particles or imperfections on optical surfaces. It is called ‘in-field’ stray light when the false light originates from within the field of view (FoV) and ‘out-of-field’ when it originates from outside the FoV. SO/PHI-HRT was designed to mitigate these effects using internal light baffles and field stops \citep[see Section~\ref{sec:understand} and][for further details]{2018SPIE10698E..4NG}. 

Past works to characterise the stray light of space-based solar magnetographs include \citet{2007A&A...465..291M} for the Michelson Doppler Imager on the Solar and Heliospheric Observatory \citep[SOHO/MDI,][]{1995SoPh..162..129S,1995SoPh..162....1D}, \citet{2009A&A...501L..19M} and \citet{2008A&A...487..399W} for the Solar Optical Telescope on the Hinode spacecraft \citep[Hinode/SOT,][]{2008SoPh..249..167T,2007SoPh..243....3K} and \citet{2014A&A...561A..22Y,2016SoPh..291.1887C} for Helioseismic and Magnetic Imager on the Solar Dynamics Observatory \citep[SDO/HMI,][]{2012SoPh..275..229S,2012SoPh..275..207S,2012SoPh..275....3P}. They employ a variety of analysis techniques, most common of which are solar limb profile analysis and planetary transits across the solar disk. Another technique to model contributions from stray light (but also the effects of insufficient spatial resolution) is to add a second non-magnetic component to the model atmosphere when performing radiative transfer inversions \citep[eg.][]{2007MmSAI..78..148L,2007ApJ...662L..31O,2011ApJ...731..125A}.

In Section 2 we first introduce the origins and limitations of the methods employed to mitigate stray light contributions in SO/PHI-HRT. We then estimate the stray light of SO/PHI-HRT to generate an extended PSF in Section 3 and interpret those results. In Section 4 we analyse the impact of the extended PSF on the SO/PHI-HRT data products and in Section 5 we outline an updated comparison of the inferred magnetic field data products with those from SDO/HMI. In Section 6 we summarise these results, and compare them with the first comparison made with SDO/HMI in \cite{2023A&A...673A..31S}. 

\section{Understanding stray light in SO/PHI-HRT}\label{sec:understand}

\subsection{Origin and suppression limitations of the SO/PHI-HRT false light contributions}

Both, the origin and the suppression of false light in solar telescopes is rather different compared to night time observatories. Night time telescopes must be efficiently protected from bright objects, which could illuminate the optical system. These objects, such as the moon, can be far away from the science field-of-view. The whole field, from which light can be brought into the instrument is called `illuminating field' and can cover as much as a full hemisphere. In order to suppress stray light contamination, it is beneficial to prevent for example the moonlight from reaching the telescope optics and mechanical structure. The most efficient way of achieving this is by an aperture baffle, or – in ground-based astronomy - by the observatory dome, which limits the acceptance angle with respect to the illuminating field.  When the optical system or its mechanical structure is illuminated by light from bright objects, many different mechanisms lead to contamination of the image with false light. 

In general nomenclature often the term `stray light' is used, regardless of the physical mechanism of light redistribution, which is not necessarily related to `scattering'. Physically, the term `scattering' of light is reserved for the interaction of light with objects, which are comparable to the wavelength of the light within a few orders of magnitude. Scattering therefore produces angular distributions of the scattered photons, which are quite broad, covering up to 4$\pi$ steradians, with more or less peaked distribution functions. Stray light modelling uses assumptions on the distribution functions in order to predict how many photons from outside the FoV are scattered into the science FoV. 

In the observation of the solar photosphere, the only disturbing light source is the Sun itself. This means that the illuminating field and the science field of view are at least partially coinciding; in a full disc telescope, they are per definition identical. The size of the solar disc is very small as compared to the half sphere, and classical `scattering' does not contribute significantly to false light in the detector. Under such an observing geometry, even a fully dusty mirror does not produce relevant stray light. While a large number of photons are scattered out of the beam path, only very little light is scattered into the beam path towards the detector. The dominant physical mechanisms, which lead to a redistribution of photons within the very limited illuminating FoV in solar observations, is diffraction and optical aberrations, in combination with the fact that the Sun is a continuous bright object. 

A sunspot represents a dark feature in front of a bright background, and the full Sun will contribute to filling this dark spot on the detector by sending photons, which have `almost' the right direction, therefore falsifying the measured signals. This also greatly impacts pores and dark intergranular lanes which reduce the intensity contrast, an important metric to provide information on the thermal variations and solar convection in the photosphere \citep[e.g.][]{2010ApJ...723L.154H,2013A&A...550A..95Y,2019A&A...621A..78K,2023A&A...675A..61K}. Furthermore the intensity contrast is also particularly important for irradiance models \citep[][]{2019A&A...624A.135Y}. Bright features are also impacted by stray light, including their polarisation signal \citep[eg.][]{2014A&A...568A..13R}. 

Even for non-aberrated systems limited only by diffraction, a significant portion of light will be seen in the sunspot, or outside the solar limb, since only 88 percent of the light will be contained within the central peak of the Airy pattern, whereas the remaining 12 percent will be distributed over the diffraction rings. The main contributor to contrast loss in solar imaging is low to mid order optical aberrations. They lead not only to a broadening of the central peak, but to an enhancement of the secondary maxima, and to an increase of light of the minima, which are not zero any more. All this occurs at the expense of the energy content within the central peak, and therefore of its amplitude. Low order aberrations have spatial scales, which are of similar order as the aperture of the telescope, which determines the width of the central diffraction peak of the non-aberrated image. \cite{2008A&A...484L..17D} have investigated the quantitative effect of low order aberrations in the case of the Spectropolarimeter (SP) of the Hinode/SOT observations. 

The PSF of the telescope is usually modelled either as the non-aberrated Airy pattern, or parameterised through a wavefront fit based on a sum of (a limited number) of Zernike or Karhunen-Loeve polynomials and their relative amplitudes \citep[see][]{1976JOSA...66..207N}. The first Zernike terms are typically used to represent the wave front of an aberrated system in phase diversity reconstructions and are able to reproduce only spatially slowly-changing distortions of the wavefront. Deformations of the wavefront that change more rapidly in space cannot be correctly represented in this way. Those spatial scales, which are macroscopically large (as compared to `scatterers' like dust, scratches and digs), but are too small to be represented by Zernikes, are called intermediate or mid-spatial-frequency (MSF) errors. They are the dominating contributors to residual false light in solar images, which have been `corrected' by deconvolution with the `PSF'. Usually the false light contribution by MSF errors is modelled with a Gaussian or Lorentz distribution (or with a convolved version of the PSF with these distributions), which is typically 10-1000 times (depending on the size of the aperture) wider than the core of the diffraction-limited PSF since the spatial frequencies correspond to physical errors from mm to several centimetres in size. 

MSF errors can have different origins. Lightweight mirrors typically show a print-through of the light-weighting geometry, when they are classically polished (`pocket effect'). In order to polish lightweight optics, non-classical polishing techniques are employed, like fluid jet polishing, magnetorheological polishing (MRF), magnetorheological abrasive flow finishing (MRAFF), or ion beam figuring (IBF). All those polishing techniques have tool sizes of several millimetres, leading to mid spatial frequencies of corresponding size. Those non-classical polishing methods are also used in the polishing of aspheres, especially when they are not rotationally symmetric, or of higher order. The choice of the polishing method is always a trade-off between acceptable shape error (local instead of classical large-scale polishing) and smoothness (classical instead of local polishing). 

\subsection{Limitations in false-light suppression in the SO/PHI-HRT system} 


The HRT system of SO/PHI represents a necessary compromise between performance under demanding orbital conditions, and strong resource restrictions in volume and mass. The quite large science FoV of $\text{0.28}^{\circ}\times\text{0.28}^{\circ}$ necessitates a wide-angle optical design. The Ritchey-Chrétien telescope does not have a real intermediate focus, and thus inhibits the use of a prime field stop, which would prevent downstream optical and structural parts to be illuminated by unwanted photons, arising from the relatively large illuminating field of 2$^\circ$ (the full solar disc at perihelion), which is 35 times larger in area than the science FoV. In the HRT case, a number of unsharp baffles and vanes are used, together with a field stop in the filtergraph and a Lyot-type pupil stop between the filtergraph and the camera. This system is very effective in suppressing false light from out-of-field. For more information on the baffling architecture of HRT we here refer to \cite{2018SPIE10698E..4NG}. A stray light analysis taking into account all illuminated mechanical components and their optical surface properties has demonstrated that the false light contributions by scattering at mechanical parts can be safely neglected \citep[][]{dbt_mods_00025030}. More details are presented in Appendix~\ref{append:sltesting}.

\section{Stray light analysis}\label{sec:analysis}

\subsection{Data Selection}\label{subsec:straylightdata}

To perform the stray light analysis we used several datasets, the details of which are summarised in Table~\ref{table:1}. A particularly useful dataset was acquired with SO/PHI-HRT on the 3\textsuperscript{rd} of January 2023, when Mercury transited across the Solar disk from Solar Orbiter's point of view. At this time Solar Orbiter was at a distance of $0.949$\;au, and it was off-pointed such that the full transit, which was close to the Southern limb, was observed. For the specific observation we show in Fig.~\ref{fig:fitting}, the Mercury disk was at $
\mu=0.56$, where $\mu=cos(\theta)$ and $\theta$ is the heliocentric angle. Furthermore, to test for any stray light dependence on the solar distance, two datasets with the solar limb in the FoV during October 2023 were analysed, one near perihelion, and the latter at an intermediate distance. These datasets were processed using the standard data reduction pipeline, including dark and flat field corrections, but without partial reconstruction by a PSF \citep[][]{2022SPIE12189E..1JS,2023A&A...675A..61K}. 

\begin{table}[h!]
\caption{Data Information}                 
\label{table:1}    
\centering                        
\begin{tabular}{m{55pt} m{60pt} m{80pt}}      
\hline
\hline          
\noalign{\smallskip}
Time [UTC] & Distance [au] & Feature \\  
\noalign{\smallskip}
\hline              
\noalign{\smallskip}
   2023-01-03 06:30:05 & $0.949$ & Mercury Transit \& South Limb \\    
   2023-10-10 23:50:03 & $0.305$ & West Limb \\
   2023-10-27 12:22:02 & $0.504$ & South Limb \\
   \noalign{\smallskip}
\hline                               
\end{tabular}
\end{table}

\subsection{Derivation of the extended PSF}\label{sebsecuc:extpsfderiv}

The core of the PSF of SO/PHI-HRT is obtained from phase diversity (PD) analysis as described in \citet{2024A&A...681A..58B}. The method described therein has been improved by now employing a Limited-memory Broyden-Fletcher-Goldfarb-Shanno (L-BFGS) algorithm to minimise the merit function, instead of a singular value decomposition method previously used. We then model the extended PSF as a linear combination of the PSF from PD analysis, $\text{PSF}_{\text{PD}}$, and a Gaussian modelling the stray light. We then take a radial one dimensional cut through limb observations and the Mercury transit data, and compare an observed solar profile $l(x)$ with an idealised profile $l_{\text{ideal}}(x)$, where the Mercury limb is represented by a Heaviside function. We model the relation between the observed and idealised profiles as follows:
\begin{equation}\label{eq:limbprofspatial}
    l(x) = l_{\text{ideal}}(x) \ast \Bigg(w_{1}\text{PSF}_{\text{PD}}(x) + w_{2}g_{2}(x)\Bigg),
\end{equation}
where $x$ is the distance along the one dimensional cut, $w_{1},w_{2}$ are weights and $g_{2}$ is a normalised Gaussian function. The $\ast$ operator denotes a convolution. In Fourier space this becomes:
\begin{equation}
    L(k) = L_{\text{ideal}}(k) \cdot \Bigg(w_{1}\text{MTF}_{\text{PD}}(k) + w_{2}\exp(-2\pi^{2}\sigma^{2}_{2}k^{2})\Bigg),
\end{equation}
where $L,L_{\text{ideal}}$ and the exponential term are the Fourier transforms of the quantities denoted by the corresponding lower case symbols in Eq.~\ref{eq:limbprofspatial} and $k$ is the spatial frequency. The MTF is the modulation transfer function, which is the absolute value of the complex valued OTF (optical transfer function). The OTF is the 2D Fourier transform of the PSF.

\begin{figure*}
    \centering
    \includegraphics[width=\textwidth]{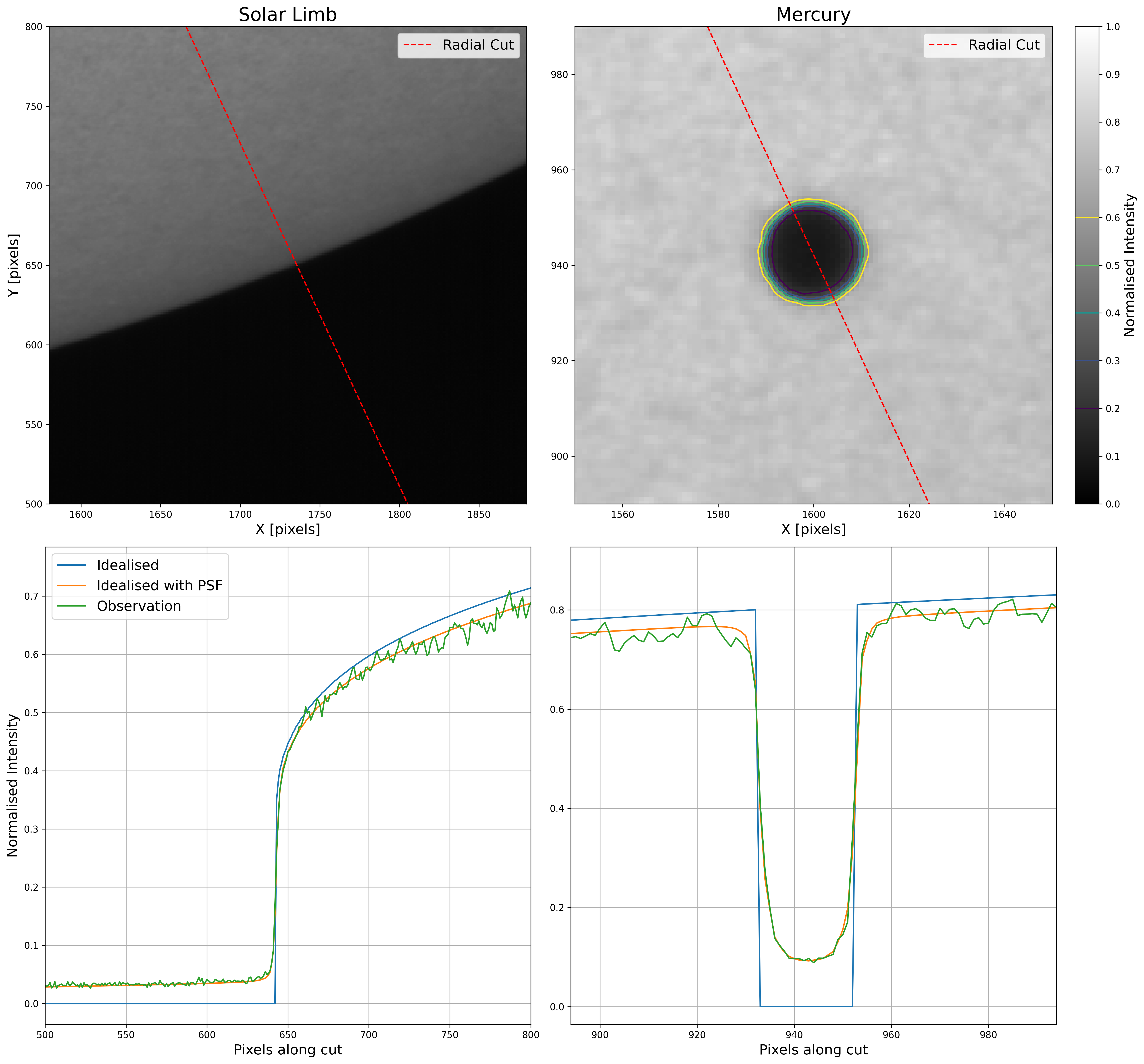}
    {\caption{Top row: A SO/PHI-HRT intensity image from 3\textsuperscript{rd} January 2023 displaying the solar limb (left) and Mercury disc (right), with a radial cut indicated by the dashed red lines. Bottom row: normalised intensities along the radial cut at the solar limb (left) and Mercury disc (right), from the observation (top row), an idealised profile and the idealised profile convolved with the best fit PSF.}\label{fig:fitting}}
\end{figure*}

The sum of the weights $w_{1} + w_{2} = 1$. The unknown or free parameters are therefore $\sigma_{2}$ and one of the weights. The idealised radial profile is generated, using a 5th order polynomial expression to describe the centre-to-limb variation \cite{1977SoPh...51...25P}. Subsequently, through trial and error the idealised profile is convolved with PSF's containing various Gaussian stray light terms until the best fit parameters are found. An example of these radial cuts is shown in Fig.~\ref{fig:fitting} for a solar limb and a Mercury transit observation. The following parameters of the Gaussian stray light term fit best for all considered datasets, independent of the distance of the spacecraft to the Sun:

\begin{equation}
\begin{split}
    w_{1} &= 1/1.11 = 0.9009... \\
    w_{2} &= 0.11/1.11 = 0.0991... \\
    \sigma_{2} &= 300"
\end{split}
\end{equation}

\begin{figure}[h!]
\centering
\includegraphics[width=\hsize]{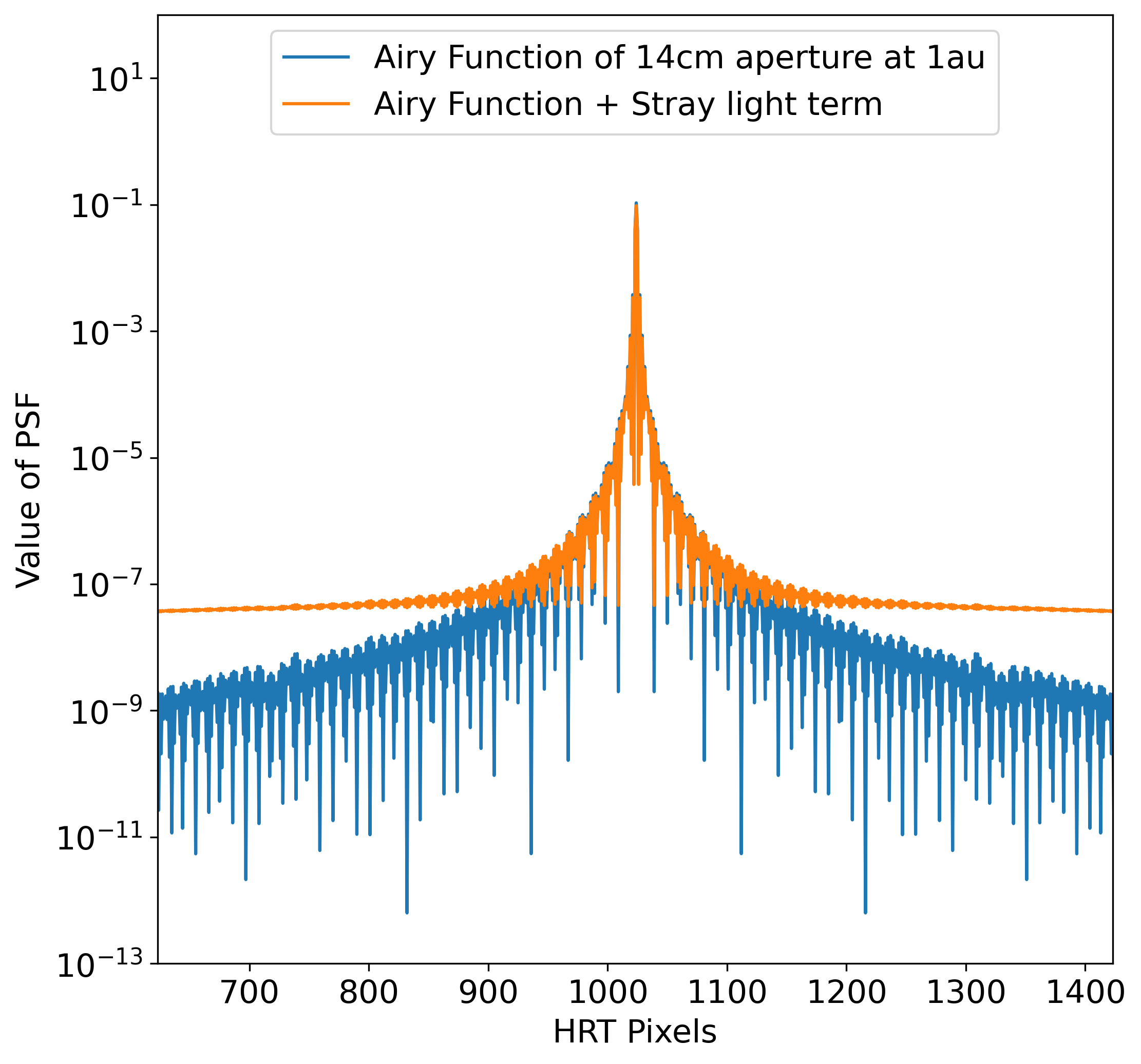}
  \caption{One dimensional cut through the centre of an Airy function for a 14 cm aperture at 1 au, and the same Airy function convolved with the best fit Gaussian stray light term. Both are normalised to have the same total energy. The Strehl ratio is $0.901$.}
     \label{fig:airy}
\end{figure}

\subsection{Interpreting the stray light analysis results}

To visually indicate the impact of the best fit stray light term, we refer to Fig.~\ref{fig:airy}. This figure indicates a radial cut through an Airy function for a $14$\;cm aperture of a perfect ideal telescope observing the Sun at $1\;$au. When the stray light term is added, one can see the wings of the PSF extending at a higher value, which at the lowest spatial frequencies in the current analysis yields about $2\times10^{-8}$, i.e. below the estimated worst case assumption (see Appendix ~\ref{append:sltesting} for the pre-flight stray light testing results). The system is still dominated by diffraction up to a radius of approximately $50$ arcseconds (100 pixels), before stray light dominates over diffraction. The angular resolution of one SO/PHI-HRT pixel is $0.5$ arcseconds. The energy content of the wide halo with respect to the central peak is about $10\%$, thus the peak amplitude is reduced accordingly. Would one give the classical Strehl estimate, one would end up with a value of $0.9$. Would one observe a star field with this telescope, the stray light halo would be orders of magnitude below the sensitivity threshold. Only the fact that the Sun is an extended bright object leads to a significant accumulated contribution in dark parts of the image.

The current analysis shows that the amount of stray light is independent of distance to the Sun. Would there be a significant contribution from out-of-field light, the `stray light' value should change, since along the orbit, the ratio of science field to illuminating field changes by a factor of almost 13. We therefore must have a closer look to the front part of the instrument, namely the entrance window and the Ritchey-Chrétien telescope. The Heat Rejecting Entrance Window (HREW) and its detrimental effect on the wave-front of the system has been discussed by \cite{2023A&A...675A..61K}. The huge thermal gradient at perihelion and the associated thermo-optical effect produces a thermal lens of up to 3 peak-to-valley waves defocus (which is compensated by the refocus mechanism in the telescope). Deviations from a pure defocus shape give rise to residual low order aberrations (mainly trefoil and spherical aberration), which have been discussed by \cite{2023A&A...675A..61K} and \cite{2024A&A...681A..58B}. They are the limiting contributors to image sharpness and granulation contrast, as they determine the shape and energy content of the PD core.  

The additional Gaussian stray light distribution of 300" width is attributed to the surface ripple of the decentred aspheres M1 and M2, which have been polished by IBF (the decentred design was chosen in order to avoid any central obstruction and spider, both, for thermal reasons and to maximize contrast transfer), see Appendix~\ref{append:sltesting} for further details. The negative effect of obstruction and spider on contrast in solar granulation has been discussed in e.g. \cite{2008A&A...484L..17D}. The decentred design also allows fully suppressing ghosting and fringing within the filtergraph \citep[][]{2018SPIE10698E..4NG}. 

\section{Application and impact of the extended PSF}


\begin{figure*}
    \centering
    \includegraphics[width=0.95\textwidth]{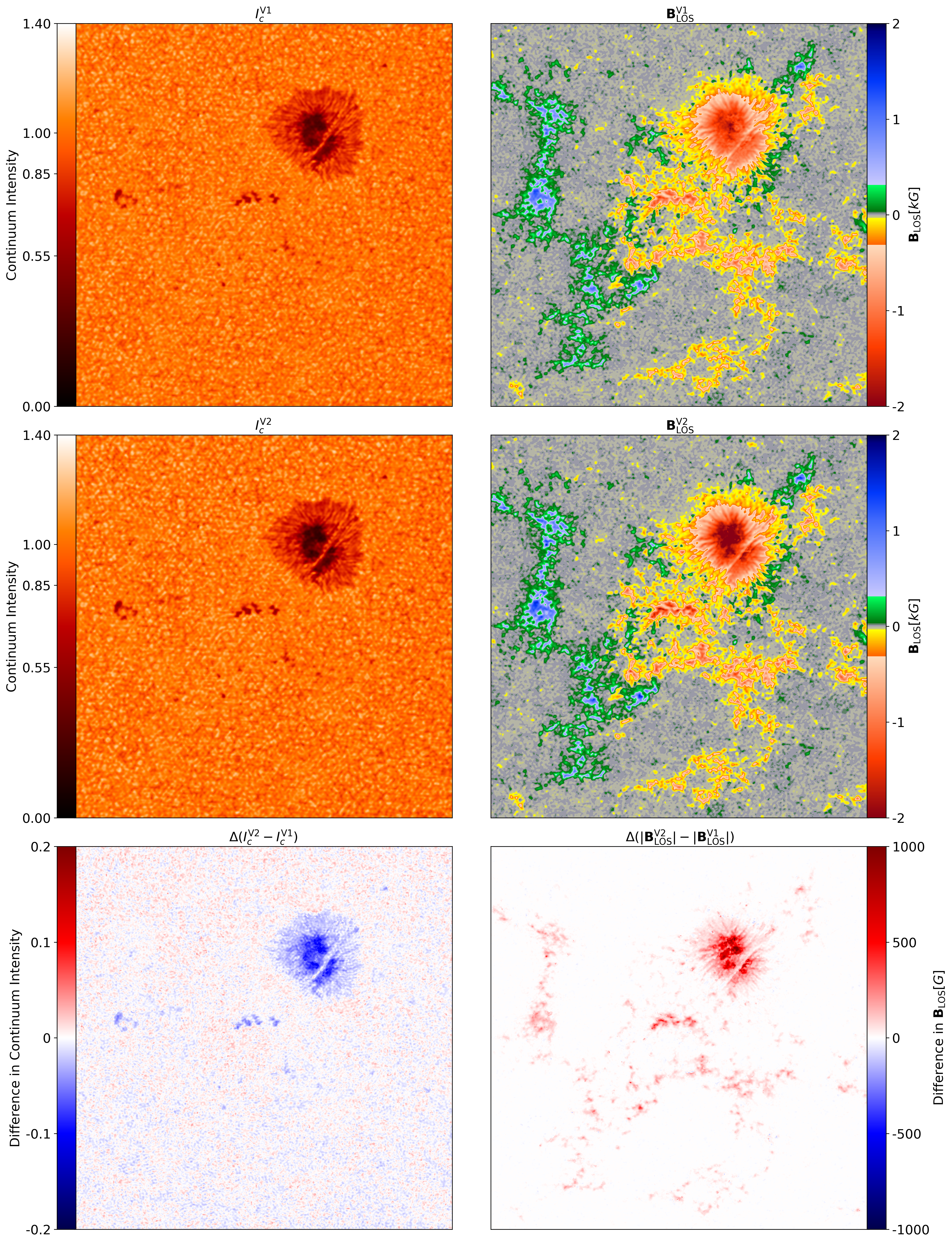}
    {\caption{Continuum intensity (left) and line-of-sight magnetic field (right) on 29\textsuperscript{th} March 2023 at 11:40 UTC. A 620 x 620 pixel portion of the FoV surrounding the sunspot and nearby plage is selected. The `V1' quantities are displayed in the top row, `V2' in the middle row, and their difference in the bottom row.} \label{fig:beforeaftericntblos}}
\end{figure*}

\begin{figure*}
\centering
\includegraphics[width=\textwidth]{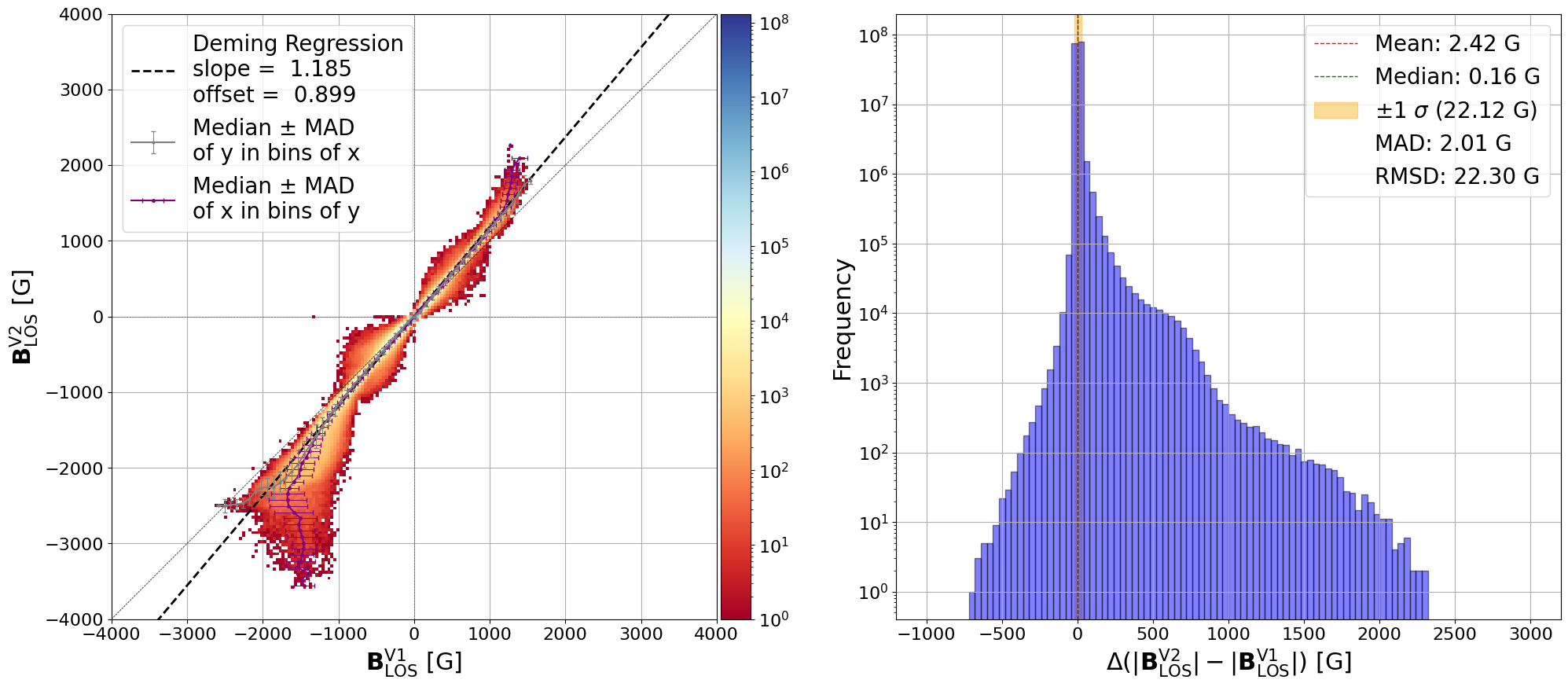}
  \caption{Left: Two-dimensional histogram comparing the `V1' and `V2' $\mathbf{B_{\text{LOS}}}$ observations from 11:40 - 13:00 UTC on 29\textsuperscript{th} March 2023. The log density of the pixels is given by the colour scale. The 1\textsuperscript{st} order Deming Regression fit and $y=x$ line are given by the dashed black and solid grey lines, respectively. The median value in bins of width $80$\;G is indicated in both $x$ (gray) and $y$ (purple) as well as the median absolute deviation (MAD) denoted by the respective error bars. Right: Histogram with bins of width $40\;$G of the difference between the absolute `V2' and `V1' $\mathbf{B_{\text{LOS}}}$. The MAD and root mean squared difference (RMSD) are indicated in the overlaid text.}
     \label{fig:blos2dv1v2}
\end{figure*}

When SO/PHI-HRT data are now produced, first the $\text{MTF}_{\text{PD}}$ is retrieved using a series of one focused and four defocused images taken on the same day with the method described in \citet{2024A&A...681A..58B}. Using the L-BFGS algorithm results in a much improved retrieval of the low to medium order Zernike terms which greatly improves the data quality (particularly at and after close approaches to the Sun). This MTF is then extended as described in Sect.~\ref{subsec:straylightdata}. The phase information from the $\text{PSF}_{\text{PD}}$ (complex component of the OTF), is then added back to the extended MTF, to form the extended OTF. The extended OTF (or PSF in real space) is then used to partially reconstruct the SO/PHI-HRT polarised Stokes data, termed  an `aberration correction' in \cite{2023A&A...675A..61K}. In this correction we reconstruct the image with a deconvolution method optimised for the phase diversity technique used in \citep[][]{1994A&AS..107..243L, 2024A&A...681A..58B} and then re-convolve with the theoretical Airy function to limit the noise increase. This partially reconstructed Stokes data is the Level 2 (L2) Stokes data that is publicly available. The L2 Stokes data then undergoes through the inversion of the radiative transfer equation in a two-pass process\citep[MILOS, see][]{2007A&A...462.1137O} to produce the L2 continuum intensity, magnetic field vector and line-of-sight (LoS) velocity. After the first inversion, the outputs are smoothed with a 15 pixel wide median filter, and used as the initial guess for the second and final inversion. For the purposes of this paper we refer to these L2 data products as `V2' (we therefore also refer to the data processed with previous versions of the pipeline as `V1') and the SO/PHI-HRT data from 2024 that are publicly released to the Solar Orbiter Archive are processed to the `V2' level. Past SO/PHI-HRT datasets from 2023 have been reprocessed and 2025 spring data at the `V2' have now been released \footnote{\url{https://www.mps.mpg.de/solar-physics/solar-orbiter-phi/data-releases}{Fourth and fifth SO/PHI-HRT data release notes}}.


Now we present the impact of applying the extended PSF to the L2 data products for a specific set of observations on 29th March 2023 from 11:40 - 15:40 UTC. At this time Solar Orbiter was in inferior conjunction with Earth, with an angular separation of $2.3$\;\degree and at a solar distance of $0.39$\;au. The high resolution instruments on Solar Orbiter during this time observed an active region (NOAA number 13262), with a sunspot at a $\mu$ value of $0.96$. During this $4$ hour observation window, SO/PHI-HRT observed on a 1 minute cadence, resulting in $240$ individual datasets. Until 13:00 UTC, the datasets were compressed with the nominal $6$-bit compression, and the remaining datasets were more strongly compressed with a $4$-bit compression scheme. For the remainder of this study, we only consider the datasets with the standard $6$-bit compression.

Further details regarding this set of observations and the impact on the LoS velocity `V2' data are presented in Calchetti et al. 2025 (under review). The L2 Stokes data, the polarised 
signal-to-noise ratio (SNR) does not significantly change due to the extended PSF. Any significant change in the SNR from `V1' to `V2' is due to the change of the improved PD analysis, not the additional stray light correction. For these observations on 29th March 2023, the improved PD analysis did not result in significant changes and therefore the SNR in Stokes $Q$,$U$ and $V$ differed by $<3\times10^{-5}I_c$. This also means that the observed changes in these `V2' SO/PHI-HRT inferred quantities that we discuss here are exclusively a result of the stray light correction alone. At or after perihelion passages, the $\text{PSF}_{\text{PD}}$ significantly changes with the improved PD method and thus so does the SNR and changes in the inferred quantities. 

First we compare the continuum intensity. In the left column of Fig.~\ref{fig:beforeaftericntblos} the `V1' and `V2' intensity images are shown for the first dataset of this set of observations on 29th March 2023. In a small patch of quiet Sun, the root mean square intensity contrast of the `V1' image is $6.45\%$ while it is enhanced to $7.01\%$ in `V2'. The quantitative differences we discuss here are representative for the entire 80 minute period of observations at the $6$-bit compression level. The difference between the `V1' and `V2' intensity images is displayed in the bottom panel of the left column in Fig.~\ref{fig:beforeaftericntblos}, and shows that `V2' has a lower intensity in the pores and sunspot, while also being brighter in the centres of granules and darker in the intergranular lanes. Quantitatively, the intensity in the dark cool umbra and nearby pores is reduced in `V2' by $15.9\%$. The mean intensity in the umbra, which we define as the area in the `V2' dataset where $I/I_c < 0.55 $, is $0.43$ and $0.37$ in the `V1' and `V2' datasets respectively.   

The changes to the Stokes profiles after stray light correction result in significantly different inferred magnetic fields after inverting the radiative transfer equations. The removal of the nearby quiet Sun intensity from the profiles in the darker features results in magnetic fields of greater strength and more vertical inclination being inferred, as the Zeeman splitting becomes more apparent. The following analysis describes only the first dataset of these observations. The mean quantities of several magnetic field parameters in the sunspot umbra and the entire FoV for the two versions of the first dataset are summarised in Table~\ref{table:2}. The mean magnetic field strength in the umbra is increased from $1657$\;G to $1994$\;G, while the field in the (negative polarity) sunspot umbra also becomes more vertical, with mean inclination changing from $143.4$\;\degree to $151.2$\;\degree. In the penumbra the mean inclination only changes by $1.4\degree$. Figures like Fig.~\ref{fig:beforeaftericntblos} for the magnetic field strength, inclination and azimuth are displayed in Appendix~\ref{append:vecdiff}. By fitting a Gaussian function to the weak-field pixels, the upper estimate of the line-of-sight (LoS) magnetic field uncertainty, $\sigma_{\text{BLOS}}$ is almost unchanged, in fact very slightly decreasing from $9.23$\;G to $8.79$\;G. Hence, the fraction of active pixels (i.e. pixels with $|\mathbf{B_{\text{LOS}}}| > 3\sigma_{\text{BLOS}}$) in the entire FoV does not significantly change, from $11.6$\% to $12.1$\%. Across the entire FoV the unsigned total LoS flux, $\Phi_{\text{USFLUX}}$, (selecting only active pixels) increases from $7.95\times 10^{21}\;$Mx to $9.05\times10^{21}\;$Mx, which, given the small change in number of active pixels, reflects the impact of the strength increase in inferred LoS magnetic fields. 

The increase in magnetic field strength, and more vertical inclination results in a significant change in the LoS fields, which is visible in the right column of Fig.~\ref{fig:beforeaftericntblos}. In the umbra the mean value increases in strength from $-1308\;$G to $-1727\;$G. A two-dimensional histogram of the LoS magnetic field for the entire 80 pairs of datasets is displayed in the left panel of Fig.~\ref{fig:blos2dv1v2}. From the median lines it is clear that on the whole stronger fields are found everywhere, but this difference dramatically increases where $|\mathbf{B_{\text{LOS}}}|>1200\;$G. On average in this regime the `V2' $|\mathbf{B_{\text{LOS}}}|$ is $253\;$G greater with a standard deviation of $153\;$G, which corresponds to a mean percentage increase of $19.7\%$ with respect to the `V1' values. The slope and offset from a first order Deming Regression are $1.19$ and $0.90\;$G. In the right panel of Fig.~\ref{fig:blos2dv1v2} the difference of the `V1' and `V2' $|\mathbf{B_{\text{LOS}}}|$ is displayed in a histogram for the entire $80$ datasets. One notices a clear positive skew with larger absolute values found across most pixels. 

\begin{table}[h!]
\caption{Mean quantities deduced from SO/PHI-HRT data at\newline 2023-03-29 11:40 UTC}                 
\label{table:2}    
\centering                        
\begin{tabular}{p{80pt} p{40pt} p{40pt} p{40pt}}      
\hline
\hline          
\noalign{\smallskip}
Quantity & V1 & V2 & \% diff. \\  
\noalign{\smallskip}
\hline              
\noalign{\smallskip}
   $I_c$ (Umbra) & $0.44$ & $0.37$ & -16\%\\    
   $\left|\mathbf{B}\right|$ [G]  (Umbra) & $1659$ & $1994$ & 20\%\\
   $\mathbf{\gamma}$ [\degree]  (Umbra)& $143.4$ & $151.2$ & 5\%\\
   $\mathbf{B_{\text{LOS}}}$ [G] (Umbra)& $-1309$ & $-1727$ & 31\%\\
   $\Phi_{\text{USFLUX}}$ [$10^{21}$Mx] & $7.95$ & $9.06$ & 14\%\\
   $\sigma_{\text{BLOS}}$ [G] & 9.23 & 8.79 & -5\%\\
   \noalign{\smallskip}
\hline                               
\end{tabular}
\end{table}

\section{Updated comparison with SDO/HMI}

 \begin{table*}
 \centering
      \caption[]{Observation details of used SO/PHI-HRT and SDO/HMI  data.}
         \label{data_table}
         \begin{tabular}{lrccccc}
            \hline
            \hline
            \noalign{\smallskip}
             & \multicolumn{2}{c}{SO/PHI-HRT} & \multicolumn{4}{c}{SDO/HMI} \\
            \noalign{\smallskip}
            \hline
            \noalign{\smallskip}
            Start time & \multicolumn{2}{c}{2023-03-29 11:40:09 \text{UTC}} & \multicolumn{4}{c}{2022-03-29 11:45:00 \text{TAI}} \\
            End time & \multicolumn{2}{c}{2023-03-29 13:00:09 \text{UTC}} & \multicolumn{4}{c}{2023-03-29 13:06:00 \text{TAI}}        \\
            Distance & \multicolumn{2}{c}{$0.394 - 0.393$ au} & \multicolumn{4}{c}{$0.998$ au} \\
            ISS mode & \multicolumn{2}{c}{On} & \multicolumn{4}{c}{On}\\
            Processing & \multicolumn{2}{c}{Ground} & \multicolumn{4}{c}{Ground}\\
            RTE Inversion & \multicolumn{2}{c}{C-MILOS} & \multicolumn{4}{c}{VFISV} \\
            \noalign{\smallskip}
             \hline
            \noalign{\smallskip}
             & \multicolumn{2}{c}{\text{Vector}} &  \multicolumn{2}{c}{\text{Line-of-sight}} & \multicolumn{2}{c}{\text{Vector}} \\
            \noalign{\smallskip}
            \hline
            \noalign{\smallskip}
            Cadence & \multicolumn{1}{r}{$60$\,s} & $720$\,s & $45$\,s & $720$\,s & $90$\,s & $720$\,s   \\
            Number of datasets &  \multicolumn{1}{r}{$80$} & $6$ & $80$ & $6$ & $55$ & $6$ \\
            \noalign{\smallskip}
            \hline
         \end{tabular}
  \end{table*}

We now present an updated comparison of the inferred magnetic field from SO/PHI-HRT (with the stray light correction) with that from SDO/HMI, a magnetograph like SO/PHI-HRT that orbits the Earth and provides near continuous full disk Solar observations. In \cite{2023A&A...673A..31S} a first comparison was made, using the `V1' data from March 2022 and reported very similar LoS magnetic fields, while differences were noted between the vector magnetic field components. The observational data used here, recorded during inferior conjunction in March 2023, has remarkably similar characteristics to the March 2022 inferior conjunction data used in \cite{2023A&A...673A..31S}. The March 2022 data included approximately an hour's worth of co-observations with a short interruption, while here we have 80 minutes of co-observations without interruption. In both cases an active region with a negative polarity sunspot and plage was present in the co-spatial FoV. The angle between the two spacecraft at the time of observations in March 2023 was $2.3\degree$, to within a degree of the $3\degree$ angular separation during the March 2022 inferior conjunction data.

There are some subtle differences: in March 2023 Solar Orbiter was closer to the Sun, $0.39$ au, as opposed to near $0.5$ au in March 2022. Furthermore in March 2022 the image stabilisation system (ISS) of SO/PHI-HRT was not operating, while it was during March 2023. An operating ISS helps to increase the SNR and to reduce the cross-talk. While the closer distance of approach does increase image degradation due to the aforementioned aberrations from the instrument entrance window, due to our enhanced PD analysis, the noise level of the Stokes products in March 2023 is  lower than that of March 2022 ($1.3 \times 10^{-3} V/I_c$ vs.\  $1.8 \times 10^{-3} V/I_c$). In March 2022, the data was taken just after 00:00 UTC, while the March 2023 data was recorded around 12:00 UTC. Due to its 24 hour orbital period, this meant that for both datasets SDO had a large radial velocity relative to the Sun, which greatly impacts the inferred fields \citep[see][]{2016SoPh..291.1887C}, of similar magnitude but in the opposite direction (approximately $-2\;$km~s\textsuperscript{-1} instead of $+3.3\;$km s\textsuperscript{-1} in March 2022). The impact of this on the SDO/HMI data and the resulting comparison is described in the following sections.

Like SO/PHI-HRT, SDO/HMI infers the photospheric vector magnetic field and LoS components from the $6173$\,\AA\ absorption line, but at different cadences (90 s \& 720 s for the vector and 45 s \& 720 s for the LoS component respectively) and different spatial resolution. At $0.39$ au, the spatial resolution of SO/PHI-HRT is $280\;$ km, while for SDO/HMI it is $725$\;km. A brief comparison of the instruments' properties can be found in Table 1 of \cite{2023A&A...673A..31S}. 
For more information regarding the description of the SDO/HMI data and its treatment we refer the reader to Section 2.2 of \cite{2023A&A...673A..31S} and the references therein. Table~\ref{data_table} briefly summarises the observation details of the SO/PHI-HRT and SDO/HMI data employed for this updated comparison. The inversion scheme used by SDO/HMI is the Very Fast Inversion of the Stokes Vector code \citep[VFISV,][]{2011SoPh..273..267B}.

Throughout this comparison with SDO/HMI we will compare now our new standard `V2' inferred magnetic fields with the standard magnetic field data products from SDO/HMI: namely the $45$s and $720$s LoS magnetic fields and the $720$s vector magnetic fields. Additionally the SDO/HMI team kindly provided the $90$s vector magnetic field data, which are used to build up the $720$s vector maps, and are most similar to the standard SO/PHI-HRT vector data in terms of SNR and cadence. SDO/HMI also produces \verb|hmi.dcon| and \verb|hmi.dconS| data series upon request, which are complete restored images using the full known PSF that includes a stray light correction \citep[see][for more information.]{2025arXiv251113348N}. The `\_dcon' series are the line-of-sight products, where the deconvolution is applied to the filtergrams, while the `\_dconS' have the deconvolution applied to the Stokes images. Given the deconvolution is done using the full PSF, these data series are not directly equivalent as the SO/PHI-HRT `V2' data is only a partial image reconstruction. In Appendix~\ref{append:dcondonS} a short comparison between `V2' SO/PHI-HRT data and the `\_dcon' and `\_dconS' series is presented.

\subsection{Method}\label{ssec:method}
The remapping and alignment procedure is almost identical to that presented in Sect. 3 of \cite{2025arXiv251025515C}. We briefly summarise the steps here. First the World Coordinate System (WCS) of SO/PHI-HRT is updated by performing a cross-correlation of the continuum intensity maps of SO/PHI-HRT and SDO/HMI, with both in the SO/PHI-HRT detector frame. Secondly geometrical distortions are compensated by comparing sub regions in the FoV, when both SO/PHI-HRT and SDO/HMI LoS magnetic field maps are in the SDO/HMI detector frame. Thirdly all the SO/PHI-HRT data products, except for the azimuth, are spatially degraded with the theoretical ideal point spread function of SDO/HMI (Airy function of a $14$\;cm aperture at 1 au). We do not apply the spatial point spread function to the azimuth as it would otherwise greatly distort the values near the ambiguous [$0\degree$,$180\degree$] boundary and at the penumbra/quiet Sun boundary. Finally, with the WCS correction applied, the SO/PHI-HRT data products are remapped onto the SDO/HMI detector frame, including a distortion correction, using `scipy.ndimage.map\_coordinates'\footnote{\url{https://docs.scipy.org/doc/scipy/reference/generated/scipy.ndimage.map_coordinates.html}} with one single third order spine interpolation. This allows for a pixel-to-pixel comparison. When every pair of datasets from SO/PHI-HRT and SDO/HMI are compared, care is taken to ensure the difference in light travel time, and difference between UTC and TAI are both taken into account. The middle point of each dataset was taken as the reference time: `DATE-AVG' for SO/PHI-HRT and `T\_REC' for SDO/HMI. As the SDO/HMI $45$s and SO/PHI-HRT $60$s data products are at a slightly different cadence, with the SO/PHI-HRT data as the limiting factor, the closest SDO/HMI $45$s dataset to each SO/PHI-HRT $60$s dataset was found. This ensured a minimum time overlap of at least $30$s between the two datasets. The same procedure was applied when comparing the SO/PHI-HRT $60$s and SDO/HMI $90$s data products, where now the SDO/HMI $90$s cadence is the limiting factor. Here the minimum time overlap between the any two dataset pairs was $45$s.

To compare SO/PHI-HRT data with the $720$s data products from SDO/HMI, we took the mean of 12 consecutive SO/PHI-HRT Stokes datasets, each separated by $60$s, centred at the middle of the time over which the $720$s data product from SDO/HMI was integrated. Solar rotation was also taken into account. The resulting long-term averaged SO/PHI-HRT Stokes profiles were then inverted with MILOS following the nominal procedure, only updating the weights for the Stokes parameters to reflect the enhanced SNR. This procedure resulted in a Stokes $V$ noise level on the order of $5\times10^{-4}V/I_c$ and LoS magnetic field maps with uncertainties on the order of $4-5$\;G, both of which are similar levels to those reported for the $720$s data products from SDO/HMI \citep[][]{2016SoPh..291.1887C}.

\subsection{Comparison of SO/PHI-HRT and HMI LoS
magnetograms}


\begin{figure*}
    \centering
    \includegraphics[width=\textwidth]{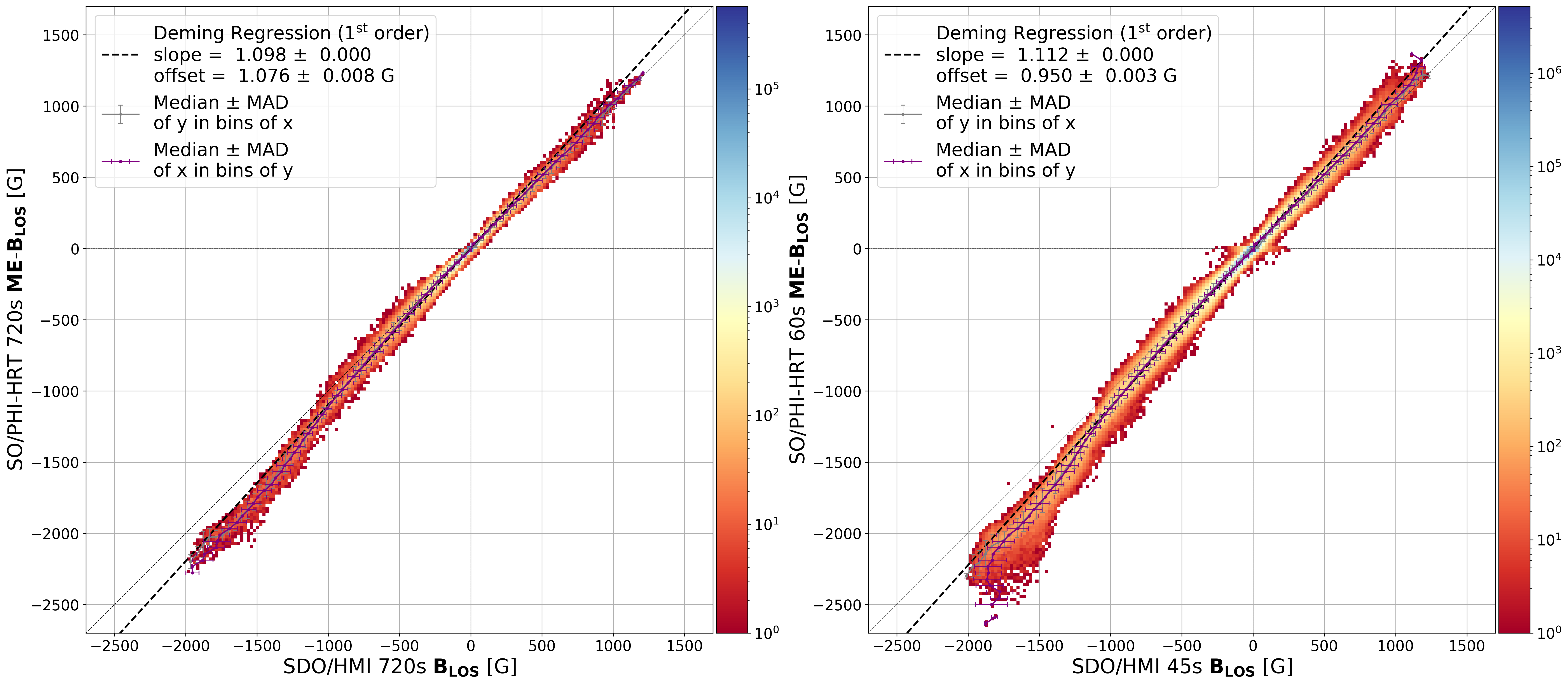}
    {\caption{Two-dimensional histograms comparing pairs of SO/PHI-HRT \textbf{ME-}$\mathbf{B_{\text{LOS}}}$ and SDO/HMI $\mathbf{B_{\text{LOS}}}$ values with $200$ bins along each axis. Left panel: SO/PHI-HRT $720$s vs SDO/HMI $720$s datasets. Right panel: SO/PHI-HRT $60$s vs SDO/HMI $45$s datasets. The log density of the pixels is given by the colour scale. The 1\textsuperscript{st} order Deming Regression fit and $y=x$ line are given by the dashed black and solid grey lines, respectively. The median value in 100 equally spaced bins is indicated in both $x$ (gray) and $y$ (purple) as well as the median absolute deviation (MAD) denoted by the respective error bars.}\label{fig:m45_m720}}
\end{figure*}

As noted in \cite{2023A&A...673A..31S}, SDO/HMI produces LoS magnetic fields via two different methods. We make a clear distinction between the LoS magnetic field computed from the inferred vector magnetic field by means of radiative transfer inversion (using the complete Stokes vector) or the LoS magnetic field computed using only the intensity and circular polarisation ($I,V$) via the MDI-like algorithm \citep[][]{2016SoPh..291.1887C}. To make this clear, from now on when referring to the LoS component of the inferred vector magnetic field from an inversion we use $\textbf{ME-}\mathbf{B_{\text{LOS}}}$ where $\mathbf{ME}$ represents a Milne-Eddington inversion. We use $\mathbf{B_{\text{LOS}}}$ to denote the SDO/HMI product produced using the MDI-like algorithm. In the left panel of Fig.~\ref{fig:m45_m720}, six pairs of $720$s $\mathbf{B_{\text{LOS}}}$ SDO/HMI (\verb|hmi.m_720s|) data and $720$s $\textbf{ME-}\mathbf{B_{\text{LOS}}}$ SO/PHI-HRT data are compared and visually represented with a two-dimensional histogram. The same is indicated in the right panel but for 80 pairs of $45$s $\mathbf{B_{\text{LOS}}}$ SDO/HMI (\verb|hmi.m_45s|) data and $60$s $\textbf{ME-}\mathbf{B_{\text{LOS}}}$ SO/PHI-HRT data. 

The slope and offset from a 1\textsuperscript{st} order Deming Regression are indicated. The Deming Regression takes into account error from both input distributions. The slope and offset of the Deming Regression, including their associated standard errors and Pearson correlation coefficient are summarised in Table~\ref{err_table} for all compared quantities in this paper. 

Both panels show very similar distributions, with near identical fit parameters and both with Pearson correlation coefficients of $0.99$. The offset is $<1\;$G, indicating parity between the data products of the zero-level. The slope values from the Deming Regression, $1.09$ and $1.07$, do not fully capture the stronger fields retrieved by SO/PHI-HRT where $\mathbf{B_{\text{LOS}}} < -1300\;$G. The mean squared differences in this strong field regime is $252\;$G and $291\;$G and a mean absolute percentage difference of $16\%$ and $19\%$ for the left and right panels respectively, using the values from SDO/HMI as the reference. Compared to the initial comparison made in \cite{2023A&A...673A..31S}, the differences noted here in this regime are larger, highlighting the stronger fields inferred by SO/PHI-HRT due to the stray light correction. 

From \cite{2016SoPh..291.1887C} a SDO radial velocity near $-2$ kms\textsuperscript{-1} results in a residual increase in $\mathbf{B_{\text{LOS}}}$ (\verb|hmi.m_|$\ast$ series) by up to $100\;$G in umbrae, with up to tens of G in penumbrae and no significant residual in the quiet Sun. Therefore the difference between the instruments may be larger when comparing SO/PHI-HRT data against SDO/HMI at other times of the day. Unfortunately, no SO/PHI-HRT data are available at other times during the conjunction.

\subsection{Comparison of SO/PHI-HRT and HMI vector magnetic
fields}\label{ssec:veccompar}
\begin{figure*}
    \centering
    \includegraphics[width=0.895\textwidth]{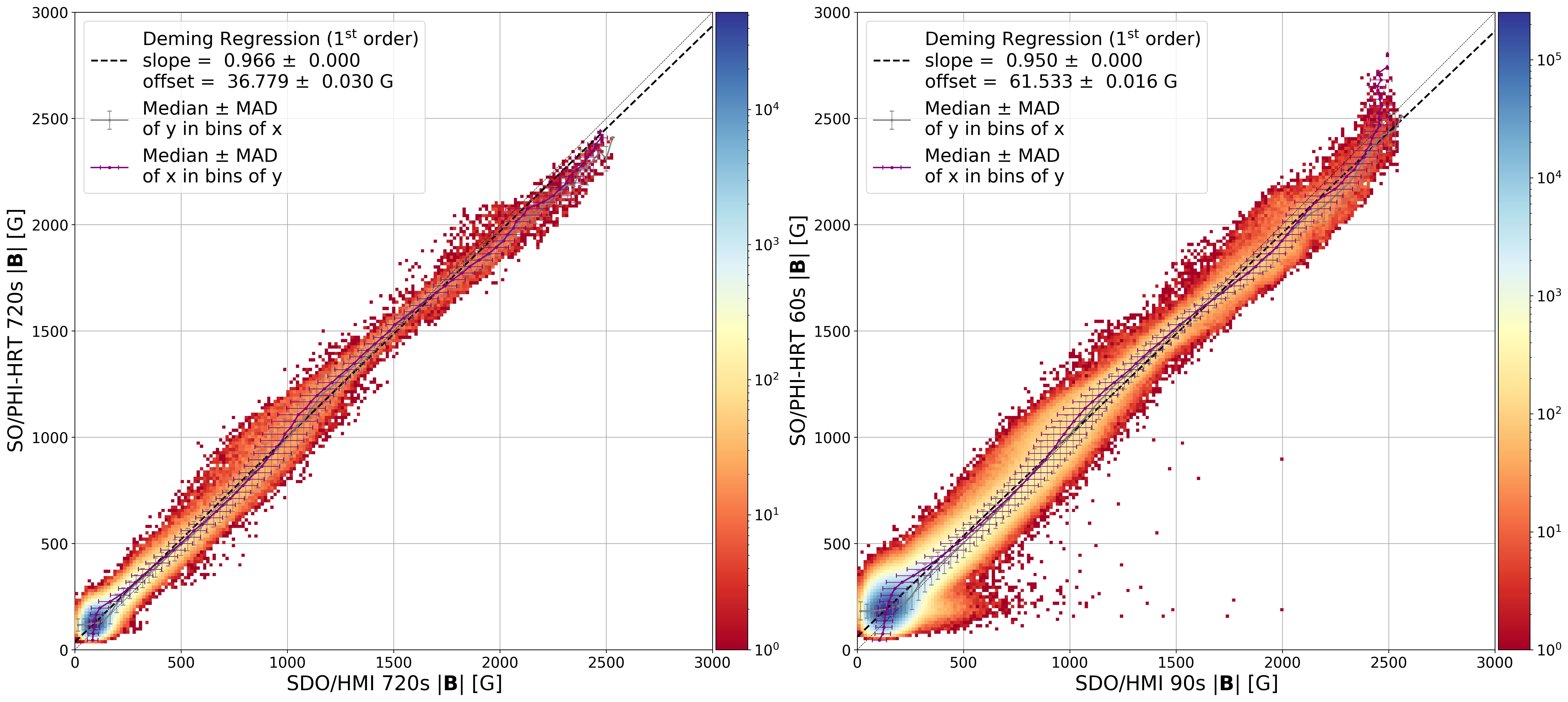}
    \includegraphics[width=0.895\textwidth]{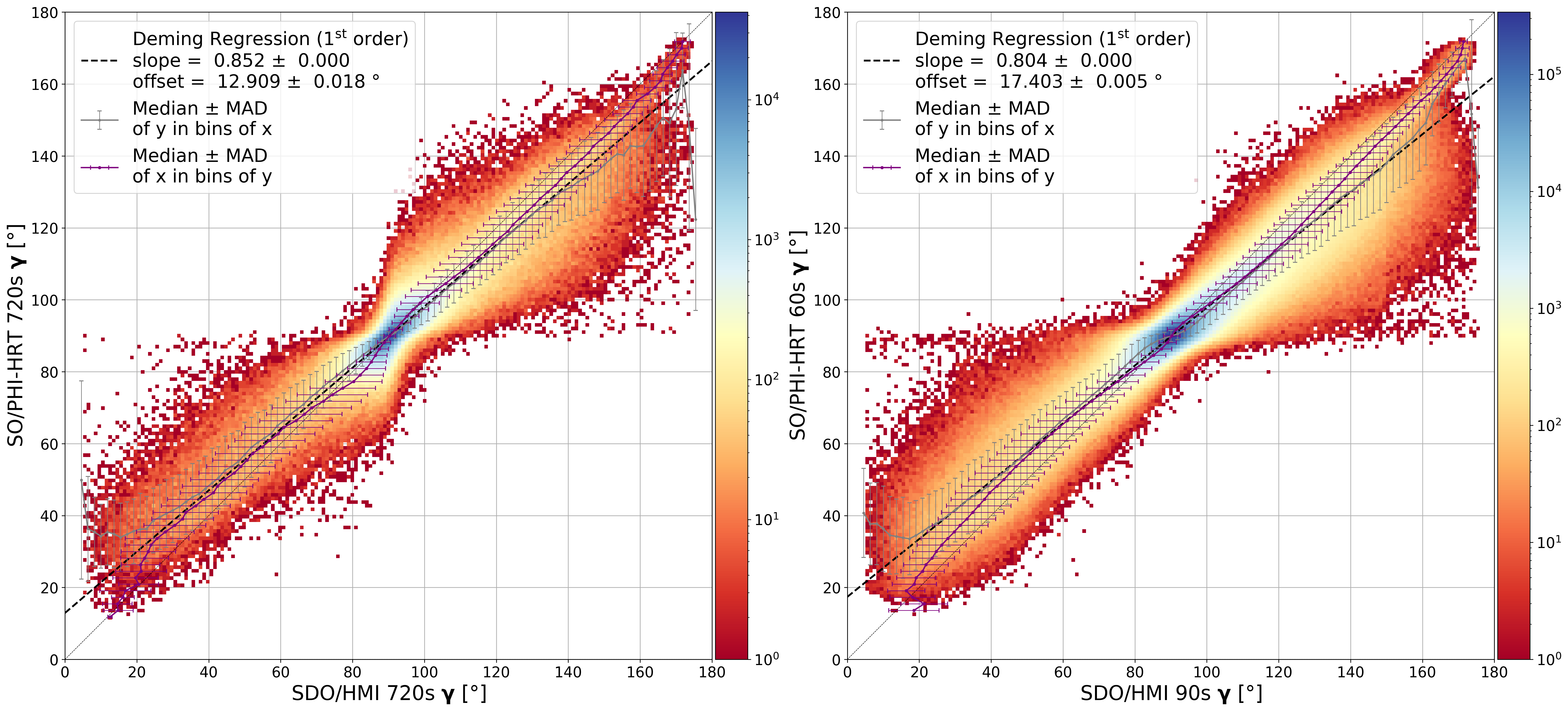}
    \includegraphics[width=0.895\textwidth]{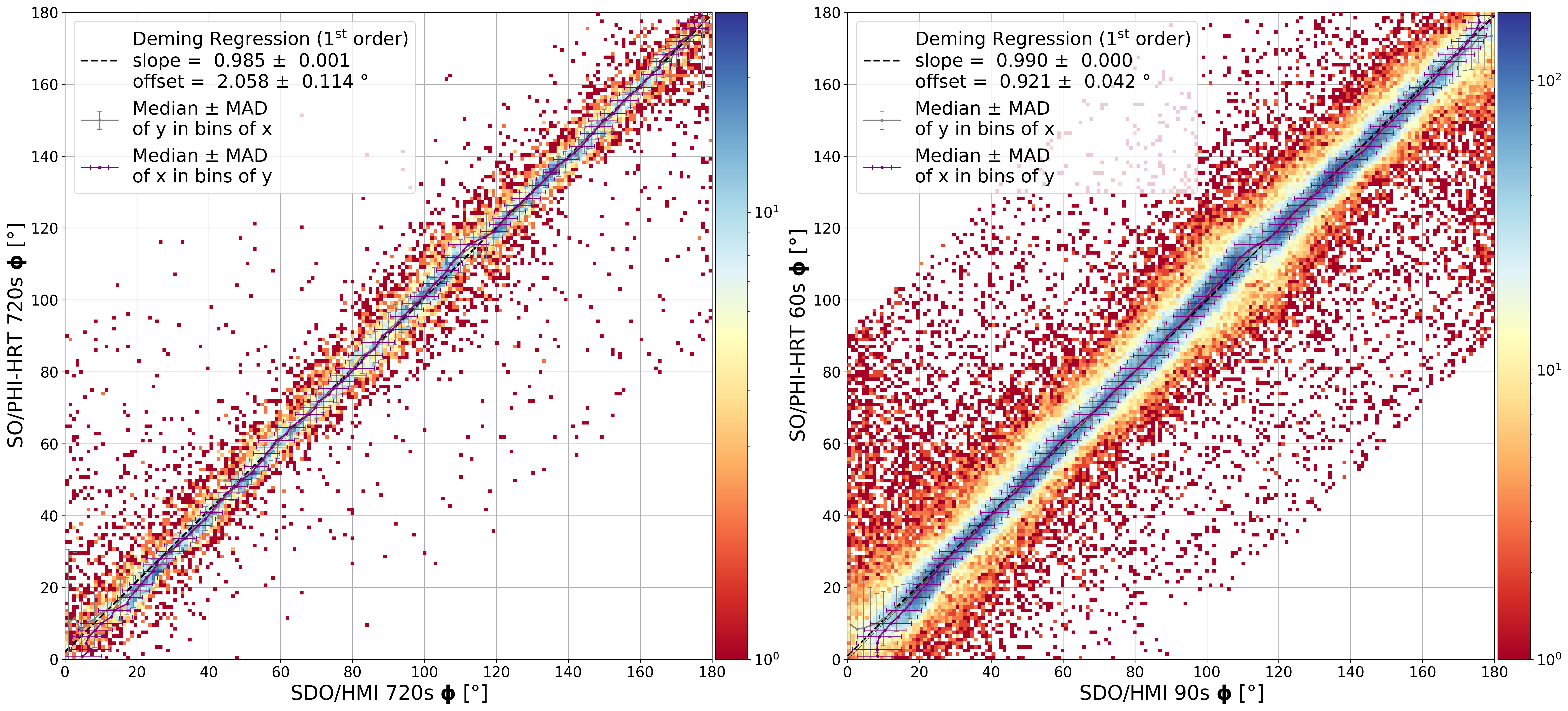}
    {\caption{Two-dimensional histograms comparing the vector magnetic field components from SO/PHI-HRT and SDO/HMI. The first column compares inversion results from $6$ pairs of SO/PHI-HRT $720$s and SDO/HMI $720$s datasets, while the second column does the same for $55$ pairs of SO/PHI-HRT $60$s and SDO/HMI $90$s datasets. Top row: magnetic field strength. Middle row: magnetic field inclination (relative to the LoS). Bottom row: magnetic field azimuth. For the azimuth, pixels where $|\mathbf{\phi_{\text{HMI}}-\phi_{\text{HRT}}}| > 90\,\degree$ or $\mathbf{\left| B\right|}_{\text{HRT}} < 600\,$G are omitted and not included in the fit. See caption of Fig.~\ref{fig:m45_m720} for details regarding the over-plotted lines.}\label{fig:veccompar}}
\end{figure*}

Now we compare the three components of the vector magnetic field, retrieved via Milne-Eddington inversions by both instruments. First we compare the magnetic field strength, $|\mathbf{B}|$. In the top row of Fig.~\ref{fig:veccompar}, one notices a good agreement between SO/PHI-HRT and SDO/HMI, particularly in the left panel, where both instrument's data products are at a $720$s cadence. The concentration of pixels near $[x,y]=[100\;$G,$100\;$G] is near symmetrical around the $y=x$ line, highlighting the similarity of the noise levels. The offsets derived from the Deming Regression fit are $36.8\;$G and $61.5\;$G, respectively. This hints that perhaps the inversion schemes of the two instruments are differently tuned for these very weak field pixels, as the noise levels of the input Stokes data are equivalent. This also points to a wider point of discussion: while both SDO/HMI and SO/PHI-HRT use Milne-Eddington inversions, they may be implemented in different ways or have different treatment of certain pixels. A comparison between VFISV \citep[][]{2011SoPh..273..267B} and MILOS \citep[][]{2007A&A...462.1137O}, similar to those already done in \cite{2014A&A...572A..54B}, is out of scope for this paper and is reserved for future work.

The slope values are $0.97$ and $0.95$, but are heavily dominated by the concentration of pixels in the weak field regime ($|\mathbf{B}|<600\;$G). In both panels the lines denoting the medians in bins of $x$ and $y$, closely follow the $y=x$ line (with a slight deviation around $1000$\;G) until a value of approximately $1600\;$G, where SO/PHI-HRT begins to infer slightly weaker magnetic field strengths. Where $|\mathbf{B}|>1600\;$G (in both SO/PHI-HRT and SDO/HMI), SDO/HMI infers stronger field strengths of $60\pm72$\;G and $56\pm78$\;G for the left and right panels respectively, where the error here denotes the standard deviation ($1\sigma$), not the standard error of the mean. The root mean squared differences in this regime are $94\;$G and $96\;$G, while the mean absolute percentage differences are $4\%$ for both cadence comparisons.

When comparing the magnetic field inclination, first we discuss the offset at the point [$90\degree,90\degree$], which is the origin of interest for the inclination in our spherical geometry coordinate system. Using the fits found from the Deming Regression, the offset at this origin is $<0.5\degree$ when comparing the $720$s data products as well as the $60$s vs. $90$s data products. When including all pixels, as we do here in the middle row of Fig.~\ref{fig:veccompar}, we naturally include many pixels where the polarisation signal is small, which results in a broad distribution of inferred inclinations by both instruments and as a consequence produces this butterfly-like spatial distribution of pixels. Like the weak field cluster of points in the magnetic field strength comparison, the concentration of pixels near [$90\degree,90\degree$] dominates the regression fit here. As in \cite{2023A&A...673A..31S} a small part of the scatter around the 1\textsuperscript{st} order Deming Regression fit is due to the angular separation between the two spacecraft, and could also result in a small offset of $<1\degree$.

When displaying these two-dimensional histograms, we have removed pixels where SDO/HMI infers $\mathbf{\gamma}>175$ or $\mathbf{\gamma}<5$. The unprocessed SO/PHI-HRT inclination maps also contain a similar fraction of pixels with very vertical fields, however the post-processing (Section~\ref{ssec:method}), specifically the application of the SDO/HMI PSF on the SO/PHI-HRT data products to reduce the spatial resolution to the SDO/HMI pixel size, mixes these vertical pixels with neighbouring inclined fields. This results in an artificial upper and lower limit in inclination in the processed SO/PHI-HRT data of approximately $15\degree$ and $165\degree$. To avoid this, the SDO/HMI PSF should be applied to the Stokes profiles from SO/PHI-HRT and then inverted, however as the wider scientific community will study and combine the standard inversion products we choose to compare them instead here in this manner. It is the same effect which creates an artificial lower limit in the magnetic field strength values in SO/PHI-HRT (visible in the top row of Fig.~\ref{fig:veccompar}). 

\begin{figure*}
    \centering
    \includegraphics[width=\textwidth]{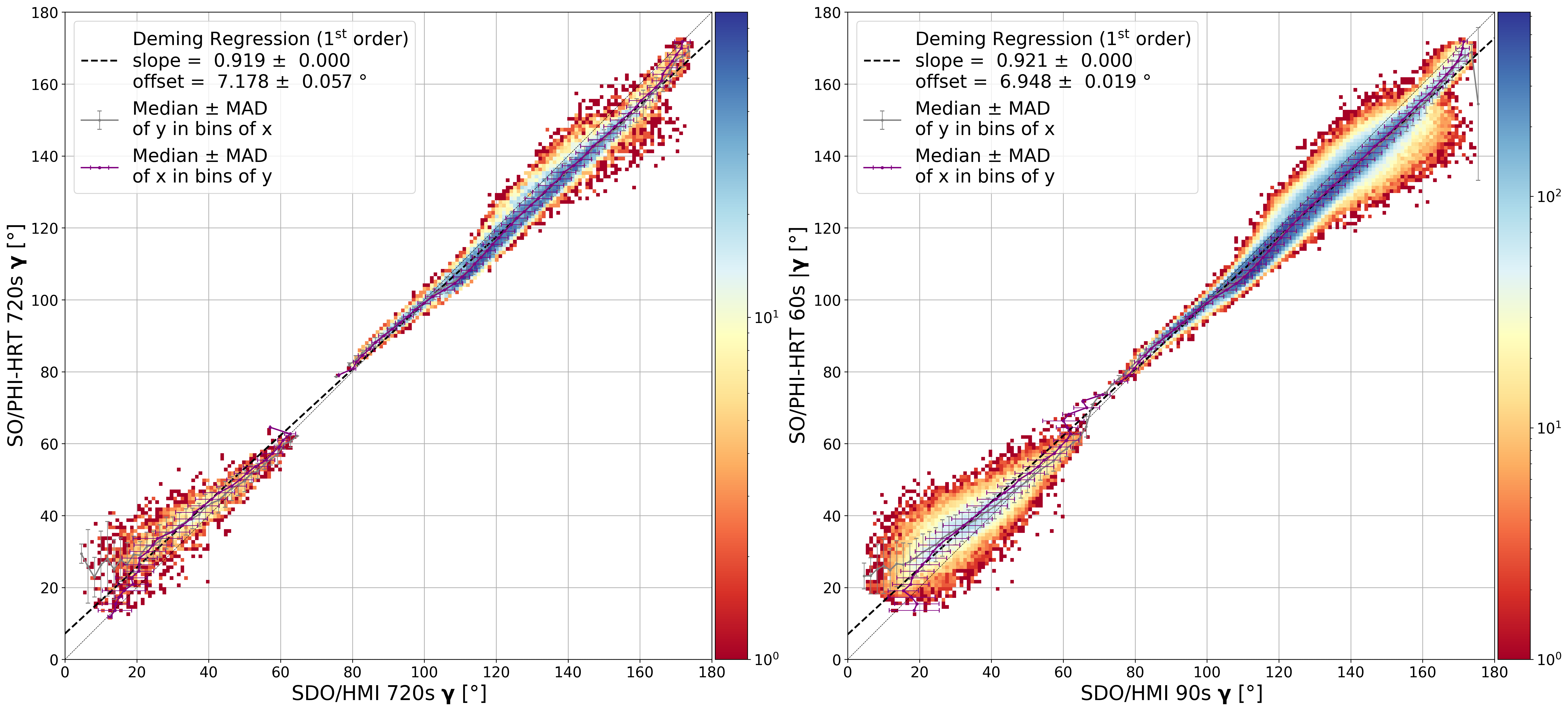}
    {\caption{Two-dimensional histograms comparing pairs of magnetic field inclination from SO/PHI-HRT and SDO/HMI $\mathbf{\gamma}$ with $180$ bins along each axis, for all pixels where $\mathbf{\left| B\right|} > 600\,$G (inferred by both SO/PHI-HRT and SDO/HMI). Left panel: SO/PHI-HRT $720$s vs SDO/HMI $720$s datasets. Right panel: SO/PHI-HRT $60$s vs SDO/HMI $90$s datasets. The log density of the pixels is given by the colour scale. The 1\textsuperscript{st} order Deming Regression fit and $y=x$ line are given by the dashed black and solid grey lines, respectively. The median value in 100 equally spaced bins is indicated in both $x$ (gray) and $y$ (purple) as well as the median absolute deviation (MAD) denoted by the respective error bars.}\label{fig:binc600g}}
    
\end{figure*}

Of greater interest is the comparison of the pixels where there is significant signal. Using $|\mathbf{B}|>600\;$G (in both SO/PHI-HRT and SDO/HMI), as a threshold to select only strong field pixels, the resulting two-dimensional histograms are depicted in Fig.~\ref{fig:binc600g}. There are a lack of points in the [$60\degree$ - $90\degree$] range, as the only sunspot in the overlapping FoV has a negative polarity, resulting in pixels with inclinations of $>90\degree$. The pixels in the range [$20\degree$ - $60\degree$] arise from the nearby positive polarity plage. In these strong field pixels SDO/HMI infers a magnetic field that is slightly more vertical, by $2\pm5\degree$ in both panels. The root mean squared difference is $5\degree$. Where $|\mathbf{B}|>600$\;G, SO/PHI-HRT infers $11\%$ and $10\%$ more inclined fields for the $720$s and SO/PHI-HRT $60$s vs SDO/HMI $90$s comparisons.

The magnetic field azimuth is compared in the two lower panels of Fig.~\ref{fig:veccompar}. Here we only depict pixels in the sunspot, where $|\mathbf{B}|_{\text{HRT}} > 600\;$G. Due to slightly different locations of the $[0\degree,180\degree]$ ambiguous boundary and the data processing described earlier, there are two clusters of pixels with very large differences in azimuth, which would severely influence the comparison. These are therefore excluded by applying an additional pixel selection criterion where the absolute difference in azimuth is $<90\degree$. This treatment is identical to that applied in \cite{2023A&A...673A..31S, 2024A&A...685A..28M}. Visually it is clearly evident that the azimuths from the two datasets are in very close agreement. As mentioned earlier the residual angular separation of the two spacecraft could result in an offset in the azimuth of $0-2\degree$. Solar north has been aligned between the two datasets to ensure the same definition is used. The over-plotted lines indicating the median value in 100 bins in both $x$ and $y$ follow the $y=x$ line closely and only deviate near the $[0\degree,180\degree]$ ambiguous boundary. 


\subsection{Comparison of SO/PHI-HRT and HMI LoS components
of the full vector magnetic field}

\begin{figure*}
    \centering
    \includegraphics[width=\textwidth]{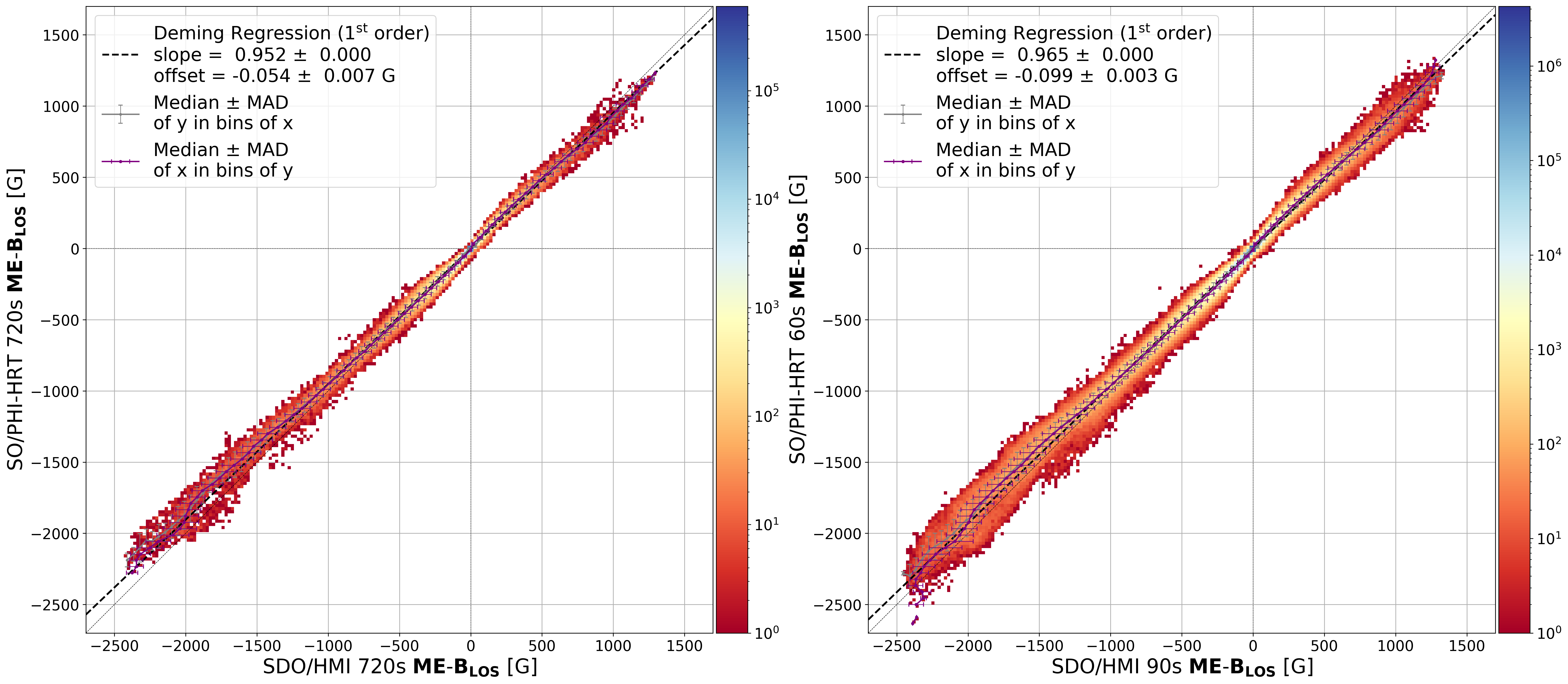}
    {\caption{Two-dimensional histograms comparing pairs of SO/PHI-HRT \textbf{ME-}$\mathbf{B_{\text{LOS}}}$ and SDO/HMI \textbf{ME-}$\mathbf{B_{\text{LOS}}}$ with $200$ bins along each axis. Left panel: SO/PHI-HRT $720$s vs SDO/HMI $720$s datasets. Right panel: SO/PHI-HRT $60$s vs SDO/HMI $90$s datasets. The log density of the pixels is given by the colour scale. The 1\textsuperscript{st} order Deming Regression fit and $y=x$ line are given by the dashed black and solid grey lines, respectively. The median value in 100 equally spaced bins is indicated in both $x$ (gray) and $y$ (purple) as well as the median absolute deviation (MAD) denoted by the respective error bars.}\label{fig:mebloscompar}}
    
\end{figure*}

In Fig.~\ref{fig:mebloscompar} we compare \textbf{ME-}$\mathbf{B_{\text{LOS}}}$ from both instruments. There is a clear trend that SDO/HMI infers stronger \mbox{\textbf{ME-}$\mathbf{B_{\text{LOS}}}$}, indicated by the slope of the Deming Regression fit: $0.95$ and $0.97$ for the left and right panels, respectively. The median in bins of $x$ and $y$ also deviates from the $y=x$ line and presents a similar picture. The root mean squared differences for all pixels is $13\;$G for both. This comparison of \textbf{ME-}$\mathbf{B_{\text{LOS}}}$ is simply another representation of the combined differences in $|\mathbf{B}|$ and $\mathbf{\gamma}$ that we obtained in the previous subsection (top two rows of Fig.~\ref{fig:veccompar}). The slightly weaker $|\mathbf{B}|$ and more inclined fields of SO/PHI-HRT result in a slightly less strong \textbf{ME-}$\mathbf{B_{\text{LOS}}}$ being inferred. Furthermore, the absolute mean difference where the SDO/HMI \textbf{ME-}$\mathbf{B_{\text{LOS}}}<-1300$ is $127\pm2$\;G and $107\pm1$\;G for the left and right panels, respectively. These differences correspond to mean SO/PHI-HRT fields $7\%$ and $6\%$ weaker, respectively. The scatter ($1\sigma$) on these difference distributions is $86$\;G and $91$\;G, respectively. The mean difference and scatter values are less than half that found in \cite{2023A&A...673A..31S} indicating a much closer agreement as a result of the stray light correction, which disproportionally affects these stronger fields.

From \cite{2016SoPh..291.1887C}, the residual of the SDO/HMI \textbf{ME-}$\mathbf{B_{\text{LOS}}}$ due to a SDO radial velocity near $-2$\;km s\textsuperscript{-1} in the umbra and penumbra is approximately $30-40$\;G, accounting for approximately $30\%$ of the noted difference found above in the strong field regime.


\begin{table*}
 \centering
      \caption[]{Quantities compared, the Deming Regression fit, absolute errors on the slope and offset, and Pearson correlation coefficient (cc)}
        \label{err_table}
        \begin{tabular}{lcccc}
        \hline
        \hline
        \noalign{\smallskip}
        Quantities compared & 1\textsuperscript{st} order Deming Regression & Slope error & Offset error & Pearson cc \\
        \noalign{\smallskip}
        \hline
        \noalign{\smallskip}
        \textbf{ME-}$\mathbf{B_{LOS}^{HRT}}$ $720$\,s vs. $\mathbf{B_{LOS}^{HMI}}$ $720$\,s & \textbf{ME-}$\mathbf{B_{LOS}^{HRT}} = 1.10 \ast \mathbf{B_{LOS}^{HMI}} + 1.1$\,G & $7\times10^{-5}$ & $0.008$ & $0.99$ \\[5pt]
        \textbf{ME-}$\mathbf{B_{LOS}^{HRT}}$ $60$\,s vs. $\mathbf{B_{LOS}^{HMI}}$ $45$\,s & \textbf{ME-}$\mathbf{B_{LOS}^{HRT}} = 1.11 \ast \mathbf{B_{LOS}^{HMI}} + 0.95$\,G & $3\times10^{-5}$ & $0.003$ & $0.99$ \\[5pt]
        $\mathbf{\left|B\right|^{HRT}}$ $720$\,s vs. $\mathbf{\left|B\right|^{HMI}}$ $720$\,s & $\mathbf{\left|B\right|^{HRT}} = 0.97 \ast \mathbf{\left|B\right|^{HMI}} + 37$\,G & $1\times10^{-4}$ & $0.03$ & $0.97$ \\[5pt]
        $\mathbf{\left|B\right|^{HRT}}$ $60$\,s vs. $\mathbf{\left|B\right|^{HMI}}$ $90$\,s & $\mathbf{\left|B\right|^{HRT}} = 0.95 \ast \mathbf{\left|B\right|^{HMI}} + 62$\,G & $7\times10^{-5}$ & $0.02$ & $0.94$ \\[5pt]
        $\mathbf{\gamma^{HRT}}$ $720$\,s vs. $\mathbf{\gamma^{HMI}}$ $720$\,s & $\mathbf{\gamma^{HRT}} = 0.85 \ast \mathbf{\gamma^{HMI}} + 13\,^{\circ}$ & $2\times10^{-4}$ & $0.02$ & $0.94$ \\[5pt]
        $\mathbf{\gamma^{HRT}}$ $60$\,s vs. $\mathbf{\gamma^{HMI}}$ $90$\,s & $\mathbf{\gamma^{HRT}} = 0.80 \ast \mathbf{\gamma^{HMI}} + 17\,^{\circ}$ & $7\times10^{-5}$ & $0.005$ & $0.94$ \\[5pt]
        $\mathbf{\phi^{HRT}}$ $720$\,s vs. $\mathbf{\phi^{HMI}}$ $720$\,s & $\mathbf{\phi^{HRT}} = 0.99 \ast \mathbf{\phi^{HMI}} + 2.1\,^{\circ}$ & $0.001$ & $0.1$ & $0.98$ \\[5pt]
        $\mathbf{\phi^{HRT}}$ $60$\,s vs. $\mathbf{\phi^{HMI}}$ $90$\,s & $\mathbf{\phi^{HRT}} = 0.99 \ast \mathbf{\phi^{HMI}} + 0.9\,^{\circ}$ & $4\times10^{-4}$ & $0.04$ & $0.97$ \\[5pt]
        $\mathbf{\gamma^{HRT}}$ $720$\,s vs. $\mathbf{\gamma^{HMI}}$ $720$\,s  (strong-field)& $\mathbf{\gamma^{HRT}} = 0.92 \ast \mathbf{\gamma^{HMI}} + 7.2\,^{\circ}$ & $5\times10^{-4}$ & $0.06$ & $0.99$ \\[5pt]
        $\mathbf{\gamma^{HRT}}$ $60$\,s vs. $\mathbf{\gamma^{HMI}}$ $90$\,s  (strong-field)& $\mathbf{\gamma^{HRT}} = 0.92 \ast \mathbf{\gamma^{HMI}} + 6.9\,^{\circ}$ & $2\times10^{-4}$ & $0.02$ & $0.99$ \\[5pt]
         \textbf{ME-}$\mathbf{B_{LOS}^{HRT}}$ $720$\,s vs. \textbf{ME-}$\mathbf{B_{LOS}^{HMI}}$ $720$\,s & \textbf{ME-}$\mathbf{B_{LOS}^{HRT}} = 0.95\, \ast \,\textbf{ME-}\mathbf{B_{LOS}^{HMI}} - 0.05$\,G & $5\times10^{-5}$ & $0.01$ & $0.99$ \\[5pt]
        \textbf{ME-}$\mathbf{B_{LOS}^{HRT}}$ 60\,s vs. \textbf{ME-}$\mathbf{B_{LOS}^{HMI}}$ 90\,s & \textbf{ME-}$\mathbf{B_{LOS}^{HRT}} = 0.97\, \ast \,\textbf{ME-}\mathbf{B_{LOS}^{HMI}} - 0.10$\,G & $2\times10^{-5}$ & $0.003$ & $0.99$ \\[5pt]
        \noalign{\smallskip}
        \hline
         \end{tabular}
  \end{table*}
  
\section{Summary}

We modelled the (high-order) stray light contribution to the PSF of the SO/PHI-HRT telescope. Using observations of a Mercury transit and limb observations, we found that a simple and broad Gaussian term modelled the impact of stray light rather well, with no dependence on the solar distance. Its independence of the distance implies that there is no significant contribution from out-of-field light. Additionally our findings are in line with pre-flight testing and modelling. 

We employed the PSF inferred through an improved version of the phase diversity technique presented in \cite{2024A&A...681A..58B} and the fitted stray light term to reconstruct the polarisation maps from the March 2023 conjunction data and found significant differences in the resulting continuum intensity and vector magnetic field data products when compared with the originally released datasets. The root mean square quiet Sun intensity contrast increased from $6.45\%$ to $7.01\%$. Intensity levels in sunspot umbrae and pores dropped by $0.07I_c$. This in turn resulted in much stronger, and more vertical magnetic fields being inferred (within a spectral line, stray light from the quiet Sun or the penumbra, by affecting $\pi$- and $\sigma$-components in different ways, is expected to have such an effect). In the sunspot umbra an $20\%$ stronger magnetic field was retrieved and the mean inclination increased by $5\%$. This combination meant that the LoS fields inferred in the umbra were $31\%$ stronger. The noise level of the magnetograms is not significantly impacted. Of greater importance for this quantity is the quality of the PD analysis and the accuracy of the retrieved PSF core. 

In SO/PHI-HRT observations at closer distances to the Sun, there is a greater difference between the `V2' and `V1' data products, as there the improved PD retrieval for the core of the PSF plays a significant role, in addition to the stray light correction.

Using very similar observational conditions, the comparison with SDO/HMI in \cite{2023A&A...673A..31S} was updated now using the `V2' SO/PHI-HRT data. On the whole a much closer agreement was found in the vector magnetic field comparison, particularly when the SO/PHI-HRT data was post-processed to the same $720$s cadence as that from SDO/HMI. With both at a $720$s cadence, the noise and weak field regions are much closer aligned. Nevertheless we remind the reader that due to the highly variable observational campaigns of Solar Orbiter and the extremely limited telemetry assigned to the instrument, SO/PHI-HRT may not always observe at a one minute cadence and therefore curating a $720$s cadence dataset post facto will not always be possible. 

Now SO/PHI-HRT infers LoS magnetic fields (by means of an inversion) stronger than the SDO/HMI fields inferred via the MDI-like algorithm, but slightly weaker than the LoS component retrieved from the complete magnetic field vector inferred via an inversion. For the $720$s comparison for $\mathbf{B_{\text{LOS}}}<-1300\;$G SO/PHI-HRT inferred LoS fields on average $13\%$ stronger than the SDO/HMI MDI-like algorithm derived $\mathbf{B_{\text{LOS}}}$, while when compared to the SDO/HMI vector derived \textbf{ME-}$\mathbf{B_{\text{LOS}}}$, inferred LoS fields on average $7\%$ weaker. Furthermore the magnetic field strengths are now closer aligned than in \cite{2023A&A...673A..31S}. While the slope values of the magnetic field inclination in the strong field pixels are similar to that found in \cite{2023A&A...673A..31S}, in the umbra the agreement is now closer.

It is now evident that the observed large scatter of the strong fields pixels in the two-dimensional histograms in \cite{2023A&A...673A..31S} were most likely the result of uncorrected stray light, which was indeed discussed as a possible cause. A potential reason discussed in \cite{2023A&A...673A..31S} for the observed differences SDO/HMI and SO/PHI-HRT are the different sampling positions across the \ion{Fe}{i} $6173.3\;\AA$\;spectral line between the two instruments (see Table~1 of \cite{2023A&A...673A..31S}). However from tests using synthetic stokes profiles and the MILOS inversion code, the different positions plays no role in the inference of the magnetic field strength, until values greater than approximately $3.5\;$kG. The different inversion schemes that SO/PHI and SDO/HMI use may play a larger role. Its study is reserved for future work. 

We would like to remind th reader that we expect these comparisons (and therefore the resulting regression fits) to vary due to multiple factors, e.g.: as a result of different viewing conditions, the SDO radial velocity and the Solar Orbiter distance to the Sun. This comparison, as that in \cite{2023A&A...673A..31S,2024A&A...685A..28M}, primarily serves to highlight that the two instruments infer very similar magnetic fields with an understanding that some relatively small residual differences do exist.

\begin{acknowledgements}
Solar Orbiter is a space mission of international collaboration between ESA and NASA, operated by ESA. We are grateful to the ESA SOC and MOC teams for their support. The German contribution to SO/PHI is funded by the BMWi through DLR and by MPG central funds. The Spanish contribution is funded by AEI/MCIN/10.13039/501100011033/ (RTI2018-096886-C5, PID2021-125325OB-C5, PCI2022-135009-2) and ERDF “A way of making Europe”; “Center of Excellence Severo Ochoa” awards to IAA-CSIC (SEV-2017-0709, CEX2021-001131-S); and a Ramón y Cajal fellowship awarded to DOS. The French contribution is funded by CNES. The SDO/HMI data are courtesy of NASA/SDO and the HMI Science Team. This project has received funding from the European Research Council (ERC) under the European Union's Horizon 2020 research and innovation programme (grant agreement No. 101097844 — project WINSUN).
\end{acknowledgements}


\begin{appendix}

\onecolumn
\section{Pre-flight stray light testing and prediction}\label{append:sltesting}

\begin{figure*}[h!]
    \centering
    \includegraphics[width=\textwidth,clip,trim=2cm 9.5cm 0cm 0cm]{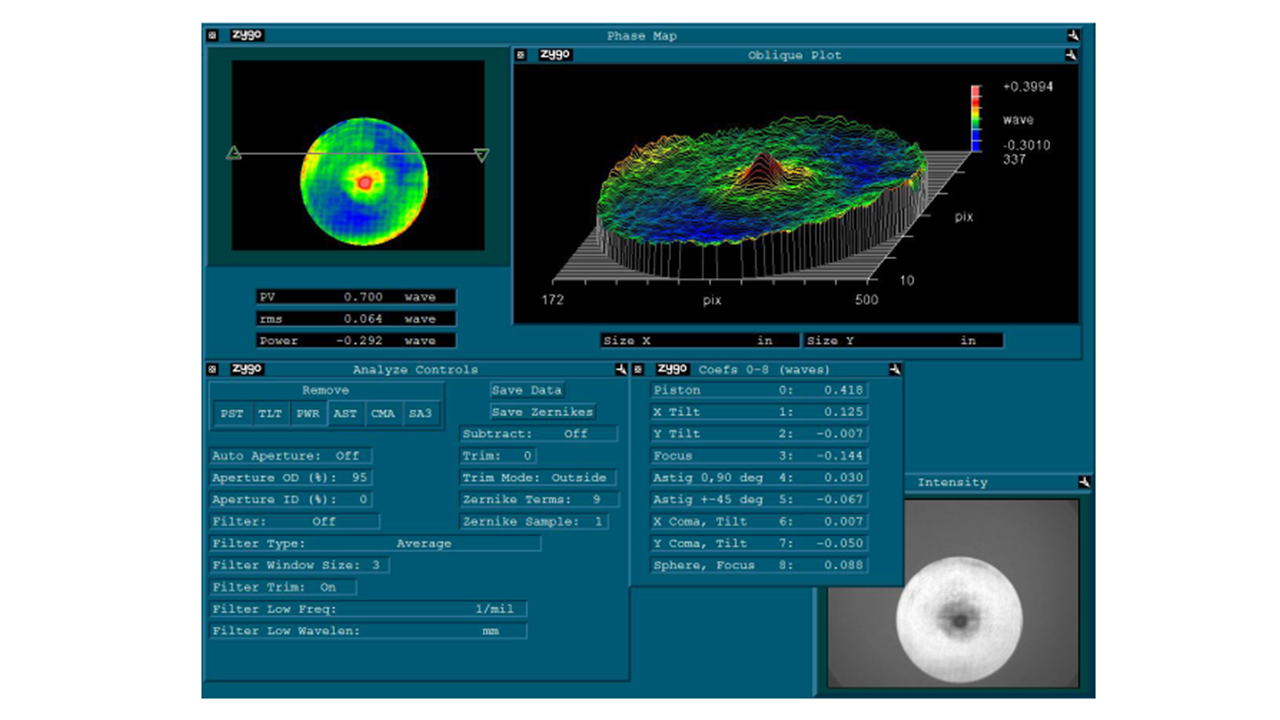}
    {\caption{Measured interferogram of the HRT telescope showing the ripple pattern of the ion beam figuring of M1. (Note that the measurement has been done in double-pass autocollimation; all wavefront values must therefore be divided by a factor of 2.)} \label{fig:interferogram}}
\end{figure*}

Fig.~\ref{fig:interferogram} shows the measured interferogram of the Ritchey-Chrétien telescope in the lab. Clearly, the path of the ion beam (the polishing tool) can be seen as a grid-like structure in the wave-front. In addition, a central peak can be seen, which is the result of a blind spot for metrology of the mirrors during the polishing procedure. For details on the mirrors of the HRT telescope we here refer to \cite{dbt_mods_00025030}. 
Simulations during the instrument development, taking into account MSF roughness of the mirrors, have given a worst case stray light background which is 7-6 orders of magnitude below the peak intensity (Bischoff and Gandorfer, 2013). 

\begin{figure*}[h!]
    \centering
    \includegraphics[width=\textwidth]{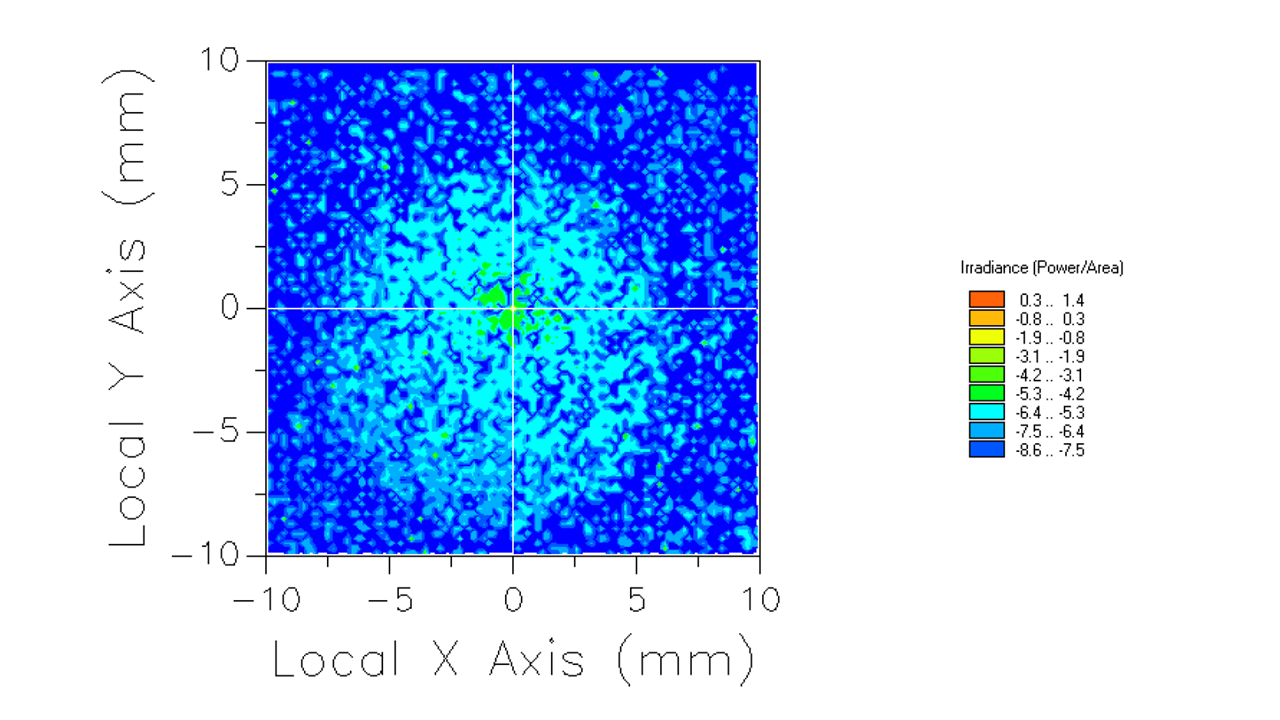}
    {\caption{Combined in- and out-of-field stray light background from the worst case straylight prediction. See text for details.} \label{fig:prediction_halo}}
\end{figure*}

\newpage

\section{Impact of extended PSF}\label{append:vecdiff}

\begin{figure*}[h!]
    \centering
    \includegraphics[width=0.95\textwidth]{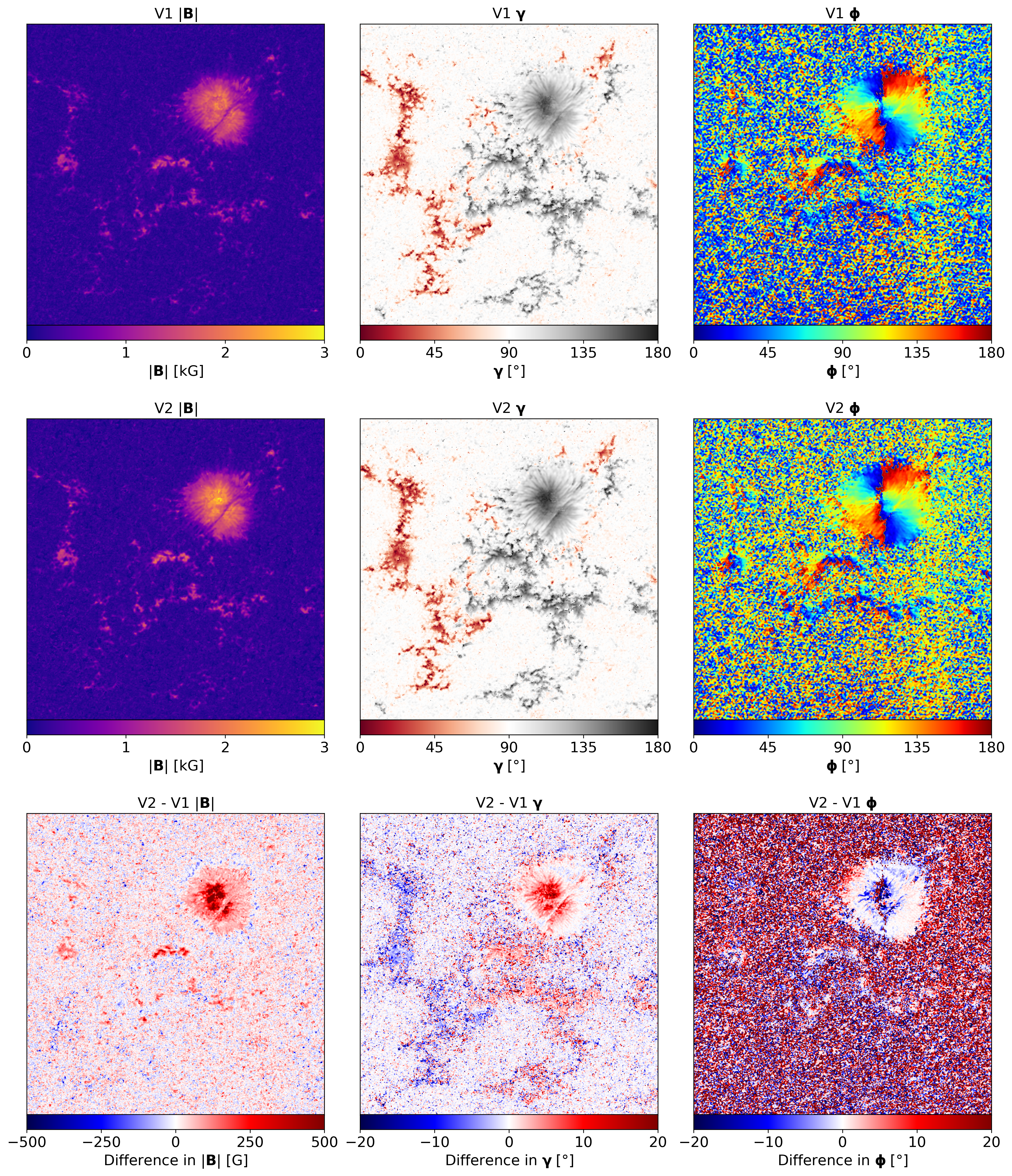}
    {\caption{Magnetic field strength (left), inclination (middle) and azimuth (right) on 29\textsuperscript{th} March 2023 at 11:40 UTC. A 620 x 620 pixel portion of the FoV surrounding the sunspot and nearby plage is selected. The `V1' quantities are displayed in the top row, `V2' in the middle row, and their difference in the bottom row.} \label{fig:beforeafterbmagbinc}}
\end{figure*}

\section{Comparison with stray light corrected SDO/HMI data}\label{append:dcondonS}

\cite{2025arXiv251113348N} describe the stray light correction for the SDO/HMI data, which was modelled in similar approach to the one described in this paper for SO/PHI-HRT. One daily stray light corrected SDO/HMI dataset is immediately available for download, but any dataset can be made available on request. As described therein, there exist two stray light corrected data series with the following appended names:`\_dcon' and `\_dconS'. The former is for the $45$s data, where the deconvolution is applied to the filtergrams, and then the derived quantities computed (without use of an inversion scheme). The latter is for the $720$s data, where the deconvolution is applied to the Stokes images, and then inverted with the VFISV Milne-Eddington code. The SDO/HMI stray light correction results in a significant increase in the continuum intensity contrast and the magnetic field strength, the latter particularly so in the plage regions by a factor between $1.4-2.5$.

However there is one implementation difference to the stray light corrected SO/PHI-HRT data we present in this work. The SDO/HMI `\_dcon' and `\_dconS' series are created through a deconvolution of the complete PSF, fully reconstructing the images. In the SO/PHI-HRT data we perform a partial reconstruction to minimise the increase in the noise levels. Therefore to put these on an equal footing and compare them in this section, we additionally convolve the stray light corrected SDO/HMI data with a theoretical PSF containing an Airy function for a 14 cm aperture at 1 au. 

\begin{figure*}[h!]
    \centering
    \includegraphics[width=0.95\textwidth]{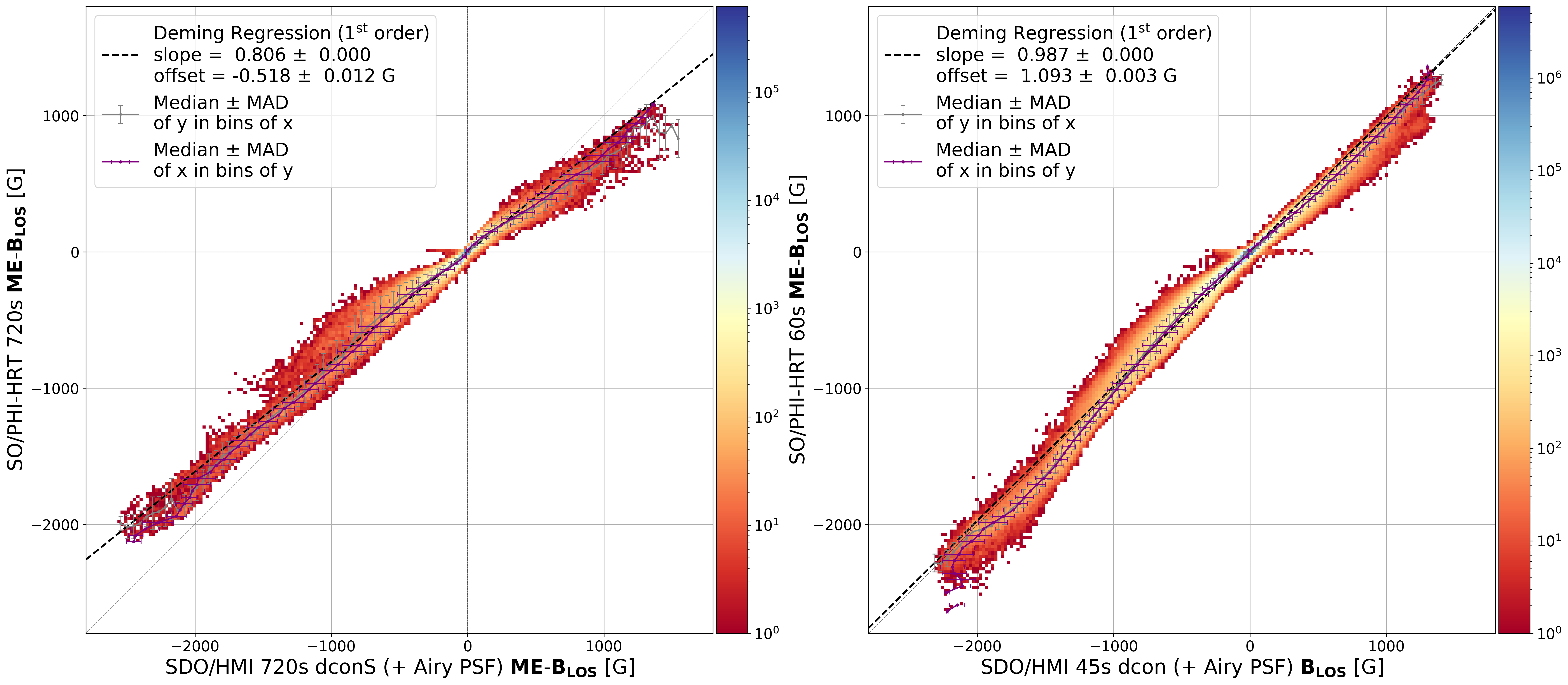}
    {\caption{Two-dimensional histograms comparing pairs of `V2' SO/PHI-HRT and stray light corrected SDO/HMI magnetograms with $200$ bins in each axis. Left panel: SO/PHI-HRT \textbf{ME-}$\mathbf{B_{\text{LOS}}}$ $720$s vs SDO/HMI \textbf{ME-}$\mathbf{B_{\text{LOS}}}$ dconS $720$s datasets. Right panel: SO/PHI-HRT \textbf{ME-}$\mathbf{B_{\text{LOS}}}$ $60$s vs SDO/HMI $\mathbf{B_{\text{LOS}}}$ `\_dcon' $45$s datasets. The `\_dcon' and `\_dconS' data from SDO/HMI is convolved with an Airy PSF (see Appendix~\ref{append:dcondonS} text for details). The log density of the pixels is given by the colour scale. The 1\textsuperscript{st} order Deming Regression fit and $y=x$ line are given by the dashed black and solid grey lines, respectively. The median value in 100 equally spaced bins is indicated in both $x$ (gray) and $y$ (purple) as well as the median absolute deviation (MAD) denoted by the respective error bars.} \label{fig:dcon}}
\end{figure*}

In Fig.~\ref{fig:dcon} the LoS magnetograms are compared: in the left panel the $720$s `\_dconS' SDO/HMI data is compared with the $720$s SO/PHI-HRT data, which both have undergone inversions. The right panel compares the $45$s `\_dcon' SDO/HMI LoS magnetic field, which is computed via a MDI-like algorithm, with $60$s SO/PHI-HRT inverted data. Both have Pearson correlation coefficients of $0.99$. As expected, the stray light corrected SDO/HMI magnetograms infer enhanced LoS fields compared to their normal data product counterparts. 


In the left panel of Fig.~\ref{fig:dcon}, the fields are much enhanced in SDO/HMI throughout the entire range, resulting in a lower slope of $0.81$ in the comparison with the SO/PHI-HRT $720$s data, as opposed to the $0.95$ found in the left panel of Fig.~\ref{fig:mebloscompar}. Here the SDO/HMI data is on average $243\pm104$\;G stronger than SO/PHI-HRT where $|B_{\text{LOS}}|>1000\;$G for both SDO/HMI and SO/PHI-HRT, compared to $101\pm86$\;G with the original $720$s data. Here the errors denote one standard deviation of the absolute differences, not the standard error on the mean.

From the right panel of Fig.~\ref{fig:dcon}, we notice that the stray light corrected SDO/HMI $45$s magnetograms are now much more closely aligned with the SO/PHI-HRT data. For stronger fields, where $|B_{\text{LOS}}|>1000\;$G for both SDO/HMI and SO/PHI-HRT, the SO/PHI-HRT $60$s data still has stronger fields, as found previously with the original $45$s data, but by a reduced amount: stronger on average by $80\pm84$\;G, compared to $119\pm72$\;G with the original $45$s data. However, the slope from the Deming Regression fit here is just below unity at $0.99$, as the SDO/HMI `\_dcon' data is very slightly stronger in the weak field regime, which dominates the regression due to the much larger number of pixels.

In an example sunspot \cite{2025arXiv251113348N} show that the stray light corrected line of sight magnetic fields, both the `\_dcon' and `\_dconS' series, are increased by a factor of $1.2$ compared to their original counterparts, which is approximately in line with the relative changes in slope we find here, compared to the slopes with the original counterparts in the main section. This however neglects the impact of the additional PSF convolution we applied to the SDO/HMI data. Without the PSF applied, the slopes will decrease further. 

\begin{figure*}[h!]
    \centering
    \includegraphics[width=0.95\textwidth]{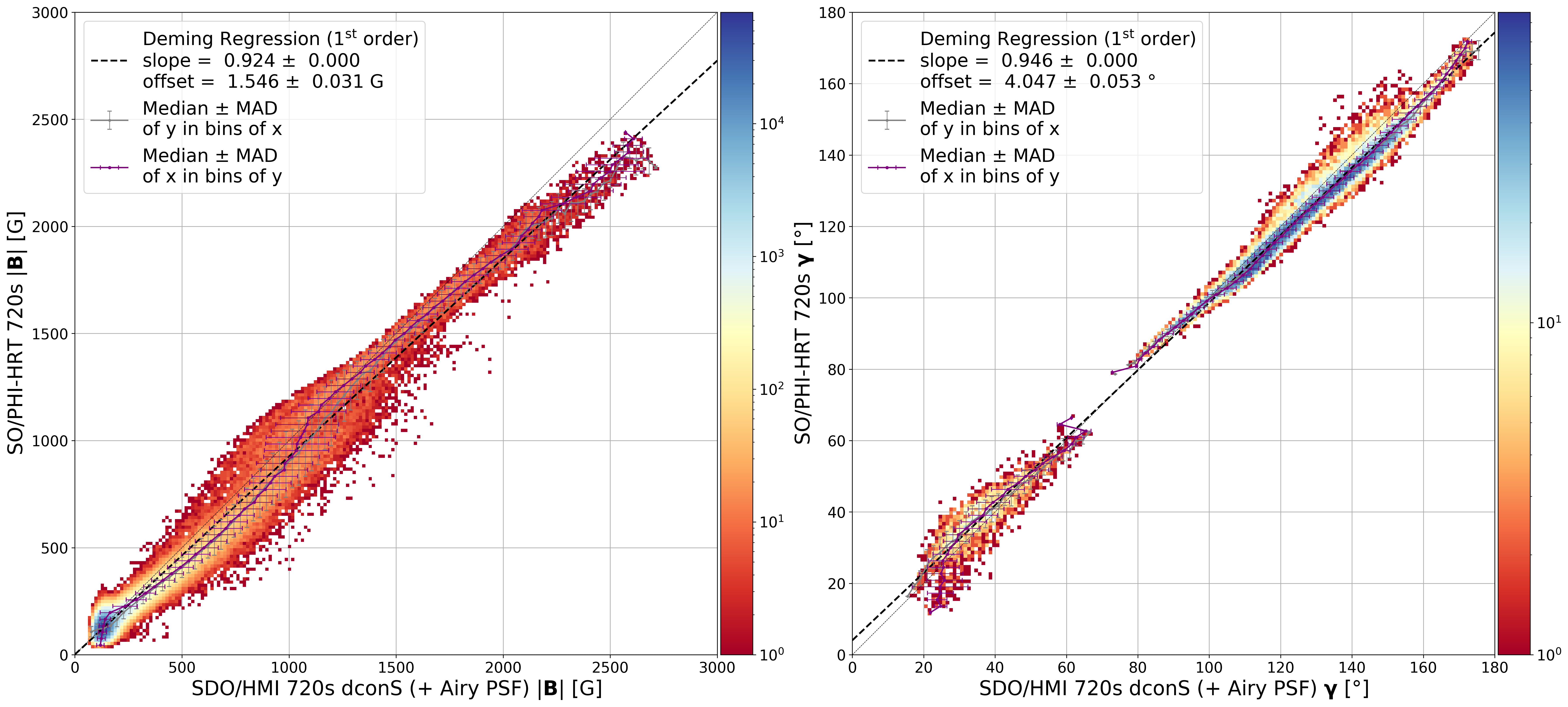}
    {\caption{Two-dimensional histograms comparing pairs of `V2' SO/PHI-HRT and stray light corrected SDO/HMI magnetic field strength and inclination. Left panel: SO/PHI-HRT $\mathbf{|B|}$ $720$s vs SDO/HMI $\mathbf{|B|}$ `\_dconS' $720$s datasets. Right panel: SO/PHI-HRT $\mathbf{\gamma}$ $720$s vs SDO/HMI $\mathbf{\gamma}$ `\_dconS' $720$s datasets where $\mathbf{|B|}>600$\;G . The `\_dconS' data from SDO/HMI is convolved with an Airy PSF (see text for details). The log density of the pixels is given by the colour scale. The 1\textsuperscript{st} order Deming Regression fit and $y=x$ line are given by the dashed black and solid grey lines, respectively. The median value in 100 equally spaced bins is indicated in both $x$ (gray) and $y$ (purple) as well as the median absolute deviation (MAD) denoted by the respective error bars.} \label{fig:dconS}}
\end{figure*}

In Fig.~\ref{fig:dconS} the magnetic field strength and inclination are compared. In general here the magnetic field strength from SDO/HMI is greater than SO/PHI-HRT for all values above approximately $250$\;G. The slope also indicates this with a value of $0.92$ as opposed to the $0.97$ value found in Fig.~\ref{fig:veccompar}. Like with the original $720$s data, SO/PHI-HRT infers weaker magnetic fields and now the mean difference, where $|B|>1600$\;G for both SDO/HMI and SO/PHI-HRT, is $124\pm83$\;G, a two-fold increase compared to the $60\pm72$\;G with the original SDO/HMI data. There is also a larger scatter of the distribution, indicated by the larger median absolute deviation (purple and grey error bars) compared to that in Fig.~\ref{fig:veccompar}. 

Finally, as described in the Section~\ref{ssec:veccompar}, we notice the consequence of the application of the SDO/HMI PSF on the lower limits of the data. Due to the application of the theoretical PSF on the SDO/HMI data to put them on equal footing, now both the SDO/HMI and SO/PHI-HRT data have artificial lower limits of approximately $40$\;G. This is why the offset between the two in the left panel of Fig.~\ref{fig:dconS}is now only $1.5$\;G, as opposed to the $37$\;G found in the top left panel of Fig.~\ref{fig:veccompar}, when the PSF was only applied to the SO/PHI-HRT data.

From the comparison of the magnetic field inclination in the right panel of Fig.~\ref{fig:veccompar} we show only the pixels where $\mathbf{|B|}>600$\;G in both SDO/HMI and SO/PHI-HRT. Here the agreement is strong, slightly more so than found in the main text, with a slope of $0.95$, and a reduced scatter around the line denoted by the median in 100 equally spaced bins.  


\end{appendix}

\end{document}